\documentclass[nofootinbib,aps,a4paper,twocolumn,preprintnumbers,11pt,superscriptaddress,showkeys]{revtex4-2}

\usepackage[left=1.3cm,right=1.3cm,top=2cm,bottom=2.3cm]{geometry}
\usepackage[utf8]{inputenc}
\usepackage[T1]{fontenc}
\usepackage[fleqn]{amsmath}
\usepackage{graphicx}
\usepackage{float}
\usepackage[font={small,it}]{caption}
\usepackage{amsmath,latexsym}
\usepackage{color}
\usepackage{hyperref}
\restylefloat{figure}
\include{amssymbol}
\input{epsf}
\definecolor{light gray}{RGB}{220,220,220}
\definecolor{dark purple}{RGB}{108,0,217}
\definecolor{pink}{RGB}{190,20,100}
\definecolor{orang}{RGB}{193,63,0}
\definecolor{green}{RGB}{11,98,17}
\definecolor{darkpink}{RGB}{153,0,76}
\definecolor{bluegreen}{RGB}{0,102,102}
\definecolor{greenlagan}{RGB}{0,102,0}
\definecolor{redgreen}{RGB}{102,102,0}
\definecolor{vividviolet}{rgb}{0.62, 0.0, 1.0}
\definecolor{amaranth}{rgb}{0.9, 0.17, 0.31}
\definecolor{palatinateblue}{rgb}{0.15, 0.23, 0.89}
\definecolor{brightpink}{rgb}{1.0, 0.0, 0.5}
\definecolor{cornflowerblue}{rgb}{0.39, 0.58, 0.93}
\definecolor{deepcarminepink}{rgb}{0.94, 0.19, 0.22}
\definecolor{radicalred}{rgb}{1.0, 0.21, 0.37}

\definecolor{beamer@PRD}{RGB}{46,48,146}
\hypersetup{
	breaklinks=true,
	pdfstartview={FitH},    
	colorlinks=true,       
	linkcolor=cyan,          
	citecolor=dark purple,        
	filecolor=magenta,      
	urlcolor=magenta,           
	anchorcolor=cyan,      
	linktocpage=true
}
\def\to{\rightarrow}
\def\d{\partial}

\def\al{\alpha} \def\be{\beta}  
   
\def\th{\theta}  
\def\ka{\kappa}
 
\def\rh{\rho}

\def\ph{\phi}

  \def\mn{{\mu\nu}}

 \def\frac#1#2{{\textstyle{{#1}\over {#2}}}}
\def\lsim{\mathrel{\rlap{\lower4pt\hbox{\hskip1pt$\sim$}}
		\raise1pt\hbox{$<$}}}
\def\gsim{\mathrel{\rlap{\lower4pt\hbox{\hskip1pt$\sim$}}
		\raise1pt\hbox{$>$}}} \def\sqr#1#2{{\vcenter{\vbox{\hrule height.#2pt
				\hbox{\vrule width.#2pt height#1pt \kern#1pt \vrule width.#2pt} \hrule
				height.#2pt}}}}

\def\beq{\begin{equation}} \def\eeq{\end{equation}}
\def\beqa{\begin{eqnarray}} \def\eeqa{\end{eqnarray}}

\begin{document}
	
	\title{Constraining the Lorentz-Violating Bumblebee Vector Field  with \\ Big Bang Nucleosynthesis and Gravitational Baryogenesis }
	
	\author{\textbf{Mohsen Khodadi}}\email{m.khodadi@hafez.shirazu.ac.ir}
	\affiliation{Department of Physics, College of Sciences, Shiraz University, Shiraz 71454, Iran}
	\affiliation{Biruni Observatory, College of Sciences, Shiraz University, Shiraz 71454, Iran}
	
	\author{\textbf{Gaetano Lambiase}}\email{lambiase@sa.infn.it}
	\affiliation{Dipartimento di Fisica ``E.R Caianiello'', Università degli Studi di Salerno, Via Giovanni Paolo II, 132 - 84084 Fisciano (SA), Italy}
	\affiliation{Istituto Nazionale di Fisica Nucleare - Gruppo Collegato di Salerno - Sezione di Napoli, Via Giovanni Paolo II, 132 - 84084 Fisciano (SA), Italy}
	
	\author{\textbf{Ahmad Sheykhi}}\email{asheykhi@shirazu.ac.ir}
	\affiliation{Department of Physics, College of Sciences, Shiraz University, Shiraz 71454, Iran}
	\affiliation{Biruni Observatory, College of Sciences, Shiraz University, Shiraz 71454, Iran}
	
	\date{\today}
	
	\begin{abstract}
		By assuming the cosmological principle i.e., an isotropic and
		homogeneous universe, we consider the cosmology of a vector-tensor
		theory of gravitation known as the \textit{bumblebee} model. In
		this model a single Lorentz-violating timelike vector field with a
		nonzero vacuum expectation value (VEV) couples to the Ricci tensor
		and scalar, as well. Taking the ansatz $B(t)\sim t^\beta$ for the	time evolution of the vector field, where $\beta$ is a free parameter, we derive the relevant dynamic	equations of the Universe. In particular, by employing observational data coming from the Big
		Bang Nucleosynthesis (BBN) and the matter-antimatter asymmetry in
		the Baryogenesis era, we impose some constraints on the VEV of the
		bumblebee timelike vector field i.e., $\xi b^2$, and the exponent
		parameter $\beta$. The former and the latter limit the size of
		Lorentz violation, and the rate of the time evolution of the
		background Lorentz-violating bumblebee field, respectively.
	\end{abstract}
	
	
	\maketitle
	
	\section{Introduction}
	The Standard Model Extension (SME), proposed by Kostelesky and
	collaborators
	\cite{Kostelecky:1990pe,Kostelecky:1994rn,Colladay:1996iz,Colladay:1998fq,Kostelecky:2003fs,Bluhm:2004ep,Kostelecky:2005ic,Kostelecky:2008ts},
	is an effective field theory that, besides describing the General
	Relativity (GR) and the Standard Model at 
	low energies, includes
	terms that violate the fundamental symmetries existent in nature,
	the Lorentz invariance, and the Charge-Parity-Time (CPT) symmetry.
	Although it is practically impossible to test these two mentioned
	symmetries at high energy due to their unavailability, the
	framework provided by SME can be used to trace them at currently
	accessible energies \footnote{As these explorations become more
		precise, some of the unknowns in the quantum gravity era (Planck
		scale) may be revealed to us. This is important since could shed
		light on the nature of the Lorentz symmetry, the same one that,
		according to the well-known approaches to quantum gravity such as
		string theory \cite{Kostelecky:1990pe,Kostelecky:1994rn},
		noncommutative field theories \cite{Carroll:2001ws}, can not stay
		invariant on any scale. }. These extra terms, introduced in the model through spontaneous symmetry breaking, address the fundamental interactions. The phenomenology of the modifications
	induced by Lorentz and CPT violating terms (see
	\cite{Mattingly:2005re,Amelino-Camelia:2008aez,Liberati:2013xla}
	for a review and references therein) has been studied in
	\cite{Kostelecky:1999zh,Yoder:2012ks,Lehnert:2003ue,Kostelecky:2001mb,Kostelecky:2002hh,Kostelecky:2006ta,Carroll:1989vb,Hohensee:2008xz,Klinkhamer:2010zs,Schreck:2011ai} for the electromagnetic sector, in
	\cite{Colladay:2009rb,Mouchrek-Santos:2016upa} for the
	electro-weak sector, and in
	\cite{Bailey:2006fd,Jacobson:2000xp,Maluf:2013nva,Maluf:2014dpa,SMEg1,SMEg2,SMEg3,SMEg4,SMEg5,SMEg6,Kostelecky:2010ze,Khodadi:2020gns,Khodadi:2022pqh}
	for the gravitational sector (for applications to gravitational
	waves, see \cite{SMEGW,SMEGW1}). The spontaneous Lorentz symmetry
	breaking (SLSB) is an elegant mechanism of Lorentz violation
	which commonly takes place when a vector or tensor field obtains a
	nonzero vacuum expectation value (VEV). On the other hand, the
	implementation of the SLSB into a curved space-time via background
	vector fields led to models that can be considered alternatives to
	GR, such as the Einstein-Aether theory \cite{aether} and the
	Bumblebee Gravity (BG) model \cite{BG2005Kost,BG2005Bert}. In this
	work, we shall focus on the latter model.
	
	The bumblebee model was initially proposed in
	\cite{Kostelecky:1989jw}  to provide a simple and
	more tractable scenario with respect to the SME
	\cite{Kostelecky:2010ze}. Inspired by the Higgs mechanism in the
	Standard Model of particles, this model also enjoys a mechanism of
	SLSB \cite{Seifert:2009gi,Bertolami:2005bh,Schreck} (see also
	Refs. \cite{Kostelecky:2003fs,Bailey:2006fd,Kostelecky:2010ze}).
	%
	The BG model, in essence, reveals a framework beyond GR via SLSB
	by the background vector field $B^\mu$ with a nonzero VEV. This means that the action of the
	bumblebee models is formed by the standard Einstein-Hilbert action
	plus terms depending on the vector field, characterized
	essentially by a kinetic term and a potential term. Here it is
	assumed that the potential has a non-vanishing VEV. The surprising
	property of this Lorentz-violating vector field model of gravity
	is that, unlike theoretical considerations in the absence of $U(1)$
	gauge symmetry, it does not forbid the propagation of massless vector
	modes\footnote{This strange feature is not unrelated to the name
		given to this model by Kostelecky, because despite the fact that
		theoretical studies prohibit the bumblebee from flying, it can
		nevertheless fly successfully \cite{Bluhm:2008yt}.}
	\cite{Bluhm:2008yt}. Due to the appearance of both Nambu-Goldstone
	(NG) and massive Higgs in theories with SLSB
	\cite{Bluhm:2004ep,Kostelecky:2005ic,SMEg2}, one expects to reveal
	a variety of physical relics in the presence of gravity which may
	be of interest in theoretical studies of dark energy and dark
	matter \cite{Bluhm:2008yt}. Recently, in \cite{Liang:2022hxd} was
	done an exhaustive analysis of the polarization of gravitational waves
	in the framework of the BG model. From viewpoint of the black hole
	phenomenology, also SLSB induced in the BG model results in noteworthy results; see for instance \cite{Liu:2019mls,Ding:2019mal,Chen:2020qyp,Maluf:2020kgf,Kanzi:2021cbg,Khodadi:2021owg,Khodadi:2022dff,Delhom:2022xfo,Maluf:2022knd,Jha:2022ewi,Kuang:2022xjp,Carleo:2022qlv,Khodadi:2023yiw}.
	
	Constraints on the bumblebee field (or its VEV) and the coupling
	constant between that field and the geometry from cosmological
	observations have been inferred from CMB \cite{maluf}. For an
	anisotropic universe, and taking the bumblebee field as $B_\mu =(0,
	b, 0,0)$, the bound derived in \cite{maluf} is $\xi b^2 <
	10^{-25}$, which is two orders of magnitude more stringent than the
	upper bound derived already from taking the bumblebee model into
	the astrophysical bodies i.e., $\xi b^2 \lesssim 10^{-23}$
	\cite{Paramos:2014mda}. In Ref. \cite{Casana:2017jkc}, owing to the
	implementation of the BG model (which includes a non-zero radial
	bumblebee field component) to justify the classical tests of GR
	within the allowed range of experimental data, it has been established some
	upper bounds on $\xi b^2$, being $<10^{-13}$ the most stringent.
	It would be interesting to note that recently in Ref.
	\cite{Xu:2022frb}, by setting a non-zero temporal component for the
	bumblebee vector field has been obtained a static spherical black hole solution and has been exposed to some classical tests.
	Moreover, by taking into account the time-like bumblebee field
	i.e., $B_\mu = (b, 0, 0, 0)$, the bumblebee cosmological model can
	be a potential candidate of dark energy to explain the present
	accelerated (de Sitter) phase of an isotropic and homogenous
	universe, provided that $\xi b^2 = 10^{-2}$ \cite{paramos}.
	
	An inevitable test of every extended theory of gravity is
	to determine the allowed regions of the model parameters via the
	confrontation with cosmological observations. Commonly these
	surveys are performed via data related to the early and late-times
	of the Universe. In this work, we explore the implementation of the
	bumblebee vector field into the cosmological background on the
	formation of primordial light elements, the \textit{Big Bang
		Nucleosynthesis} (BBN), as well as the matter-antimatter
	asymmetry in the Universe, known as \textit{Baryogenesis}. The
	former occurred in the early phases of the Universe evolution, between the first fractions of seconds after the Big Bang ($\sim 0.01$ sec) and a few hundred seconds after it (in this epoch the Universe was hot and dense).  BBN describes the sequence of nuclear reactions that yielded the
	synthesis of light elements	\cite{kolb,bernstein,Burles:2000ju,olivePDGroup2014}, and therefore drives the observed Universe. In general, from the physics of BBN epoch,  one may infer stringent constraints on a given cosmological model \cite{torres,Lambiase3,Asimakis:2021yct}.
	In particular, in the present paper, we shall derive the constraints on the free parameter of the Bumblebee cosmological
	model i.e., $\xi b^2$.
	
	Baryogenesis, the latter physical process under our attention in
	this paper is expected to have taken place during the early universe (before BBN) as the origin of the baryon asymmetry
	\footnote{Gravitational baryogenesis just not leads to baryon
		asymmetry but also may produce dark matter asymmetry
		\cite{allGB1}.}. It, in essence, addresses one of the unsolved
	problems of cosmology and particle physics, meaning that
	contrarily to what is expected from various considerations (the
	amount of matter (baryons and leptons) should equate the amount of
	anti-matter (anti-baryons and anti-leptons)), observations show that
	in the Universe matter dominates over anti-matter
	\cite{CKB,Riotto:1998bt,gravbar,Cline:2006ts,Saj,CDS}. It means
	that the observed baryon asymmetry must have been produced
	dynamically during the early universe because the Universe
	initial state with equal numbers of baryons and antibaryons. Sakharov was the first to establish the
	conditions (Sakharov's conditions) for the occurrence of such
	baryon asymmetry \cite{Sakh}\footnote{See also
		\cite{Dolgov:2005wf} for more details.}: 1) There must exist
	interactions that violate the baryon number (violation of the
	Baryon number), 2) Violation of the fundamental discrete
	symmetries: C and CP violation, 3) Deviation from thermal
	equilibrium.  In this regard, there are some possible physics
	mechanisms such as: GUT baryogenesis	\cite{Weinberg:1979bt,Nanopoulos:1979gx,Yoshimura:1979gy,Yoshimura:1996eh},
	Electroweak baryogenesis	\cite{Arnold:1987mh,Rubakov:1996vz,Riotto:1999yt}, and
	Leptogenesis
	\cite{Akhmedov:1998qx,Dick:1999je,Murayama:2002je,APS,Thomas:2005rs,Thomas:2006gr,Lambiase:2013haa}
	which is expected to explain baryogenesis (see also review paper
	\cite{Davidson:2008bu}). Some scenarios look for the origin of
	baryogenesis in Hawking radiation \cite{Hook:2014mla}, B mesons
	\cite{Elor:2018twp,Alonso-Alvarez:2021qfd}, primordial black holes
	\cite{Smyth:2021lkn}, minimal fundamental length
	\cite{Das:2021wxq}, and generalized uncertainty principle
	\cite{Das:2021nbq}.
	
	The CMB observation (through the  acoustic peaks) and the measurements of large-scale structures allow to infer an
	estimation of the baryon asymmetry parameter $\eta$:
	$\eta^{(CMB)}\sim (6.3\pm 0.3)\times 10^{-10}$ \cite{27GL}. Yet,
	estimations on $\eta$ can be also obtained from BBN, leading to  $\eta^{(BBN)}\sim (3.4-6.9)\times 10^{-10}$
	\cite{28GL}. These two values are compatible, although they are derived in two different eras of the Universe. Other values close
	to these two such as $\eta^{obs}\sim (8.6\pm0.1)\times 10^{-11}$,
	are also found in the literature
	\cite{ParticleDataGroup:2018ovx,AharonyShapira:2021ize}. As an
	application of the measurement of $\eta$, it can be used as one of
	the common ways to evaluate the viability of any extended
	cosmology model by modified gravity
	\cite{Lambiase:2006dq,Lambiase:2006ft,allGB2,allGB3,allGB4,allGB5,allGB6,allGB7,allGB8,allGB9}.
	
	By and large, with this idea that the background Lorentz-violating
	bumblebee field $B^\mu$ has a time-like component different from
	zero $B_\mu=(B(t), 0, 0, 0)$, with $B(t) \sim t^\beta$, throughout
	this paper, we focus on the early times of the Universe, in
	particular, BBN and Baryogenesis eras, to provide stringent
	constraints on $\xi b^2$. More exactly, the key purpose of this
	work is further shedding light on the SLSB induced in the
	bumblebee vector field model, through exposure to the
	above-mentioned early Universe scenarios.
	
	The paper is organized as follows. In the next Section, we recall
	the main topics of the bumblebee cosmological model, focusing on a
	homogeneous and isotropic universe (the Friedmann-Robertson-Walker
	(FRW) universe). Here with this idea that the time-like bumblebee
	field is time-varying, we solve the cosmological field equations.
	In Sections \ref{BBN}, \ref{Bar} and \ref{New} we use these dynamic
	equations to infer the bounds on the involved parameter(s) in the
	bumblebee cosmology model. Our conclusions are release in Section
	\ref{Con}.
	\section{The bumblebee model} \label{BM}
	The bumblebee model generalizes the standard formalism of General
	Relativity by allowing a SLSB. The latter manifests by means of a
	suitable potential with a non-vanishing VEV, which allows the
	bumblebee vector field $B_\mu$ to acquire a four-dimensional
	orientation.

	We consider the bumblebee action \cite{Kostelecky:2003fs,paramos}
	\beq\label{eqAction}
	\begin{split}
		S = & \int \sqrt{-g} \left[ \frac{1}{2\kappa} \left( R + \xi B^\mu B^\nu R_{\mu\nu} +\chi B_\alpha B^\alpha R \right) \right.- \\
		& - \left. \frac{1}{4}B^{\mu\nu} B_{\mu\nu} - V \left( B^\mu B_\mu \pm b^2 \right) + \mathcal{L}_M \right] d^4 x,
	\end{split}
	\eeq
	where $\kappa \equiv8\pi G$, while $\xi$ and $\chi$ are coupling constants with the same mass dimension $[\xi]=M^{-2}=[\chi]$. These two coupling constants, in essence, are responsible for controlling the non-minimal coupling between the Ricci curvature $R_{\mu\nu}$ and scalar Ricci $R$ with the bumblebee field $B_{\mu}$ (with mass dimension $[B^{\mu}]=M$), respectively. $B_\mn\equiv \d_\mu B_\nu - \d_\nu B_\mu$ is the field-strength tensor, $b^2 \equiv b_\mu b^\mu = \langle B_\mu B^\mu \rangle_0 \neq 0$ is the expectation value for the contracted bumblebee vector, and $\mathcal{L}_M$ is the Lagrangian density for the matter fields. The potential $V$ exhibits a minimum at $B_\mu B^\nu \pm b^2 = 0$. Concerning the significance of bumblebee potential form, it needs to recall that by setting two linear and quadratic forms for the bumblebee (timelike) potentials in the action (\ref{eqAction}), then its flat counterpart meets the bumblebee theory proposed by the Kostelecky and Samuel in Ref. \cite{Kostelecky:1989jw}. It is well-known from Ref. \cite{Bluhm:2008yt} that the Hamiltonian density $\mathcal{H}$ just in some very restricted region of classical phase space in Kostelecky and Samuel's model can be positive, meaning that the relevant bumblebee theory is stable. In other words, the bumblebee theories based on these two forms of potential in most regions of phase space suffer from instability, $\mathcal{H}<0$. This is also shown for other well-known SLSB-based vector theories, such as  Aether theory \cite{Carroll:2009em}. Anyway, it is not a worrying issue for the cosmological model at hand since by keeping open the general form of potential, our analysis will rule out both linear and quadratic forms \footnote{Apart from this, the bumblebee models, including gravity, are considered effective theories likely appearing below the Planck scale from a more fundamental quantum theory of gravity. In this framework, stability is expected to be restored due to the imposition of additional constraints raised by quantum gravity effects. As a result, without having a fundamental quantum theory of gravity, one can not exactly address the final stability of bumblebee models \cite{Bluhm:2008yt}.}.
	
	The variation of Eq. (\ref{eqAction}) with respect to the metric leads to the modified Einstein equations
	\beq \label{EqEinstein}
	G_{\mu\nu} = \kappa T_{\mu\nu} +T^{(B)}_{\mu\nu}
	\eeq
	where $T_{\mu\nu}$ is the matter energy-momentum tensor for matter
	\beq
	T_{\mu\nu} = (\rho +p)u_\mu u_\nu +p g_{\mu\nu},
	\eeq
	with $\rho, p$ are the energy density and pressure of matter, respectively, $u_\mu = (1,0,0,0) $ the four-velocity of the fluid with the normalization condition $u_\mu u^\mu = -1$), and $T^{(B)}_{\mu\nu}$ is given by
	\begin{eqnarray}
		T^{(B)}_{\mu\nu} &=& \ka \bigg[ 2 V' B_\mu B_\nu- B_{\mu\al}B^\al_{\nu} - \left( V + {1 \over 4} B_{\al\be}B^{\al\be}\right) g_\mn \bigg] \nonumber \\
		& & + \xi  \bigg[ {1 \over 2} B^\al B^\be R_{\al\be} g_\mn - B_\mu B^\al R_{\al\nu} - B_\nu B^\al R_{\al\mu}  \nonumber  \\
		&  &+ {1 \over 2}  \nabla_\al \nabla_\mu (B^\al B_\nu) + {1 \over 2}  \nabla_\al \nabla_\nu (B^\al B_\mu)  \\ && - {1 \over 2}  \nabla_\al \nabla_\be( B^\al B^\be) g_\mn - {1 \over 2} \Box ( B_\mu B_\nu ) \bigg]\,, \nonumber \\
		&-& \chi [ B_\al B^\al G_\mn + R B_\mu B_\nu +  g_\mn \Box(B_\al B^\al) - \nonumber \\
		& & - \nabla_\mu \nabla_\nu (B_\al B^\al) ]\,, \nonumber 
	\end{eqnarray}
	where $V'$ denotes the derivative of the potential $V$ with respect to its argument.
	
	The trace of the modified Einstein equation (\ref{EqEinstein}) reads
	\begin{equation}\label{TraceEq}
		-R = \kappa {\cal T} + T^{(B)}
	\end{equation}
	with
	\begin{eqnarray}
		{\cal T} &=& T^\mu_{\,\,\,\, \mu} =\rho-3p\,,\label{TraceEq1} \\
		T^{(B)} &=& \kappa \left[V' B_\alpha B^\alpha-B_{\alpha\beta}B^{\alpha\beta}-2V\right]- \label{TraceEq2} \\
		& & -\xi \left[\nabla_\alpha \nabla_\beta+\frac{1}{2}g_{\alpha\beta} \Box \right]B^\alpha B^\beta - 3\chi \Box B_\alpha B^\alpha \,,\nonumber 
	\end{eqnarray}
	where $\Box=\nabla_\alpha \nabla^\alpha$ is the D'Alembert operator in curved spacetimes.
	
	The variation of Eq. (\ref{eqAction}) with respect to the bumblebee field yields its equation of motion,
	\beq\label{eqbumblebee} \nabla_\mu B^{\mu\nu} = 2   \left( V'
	B^\nu - \frac{\xi} {2\kappa}   B_\mu R^{\mu\nu} - \frac{\chi}
	{2\kappa}   B^\nu R  \right)\,. \eeq
	If the LHS of the equation
	vanishes, the above results in a simple algebraic relation between
	the bumblebee, its potential and the geometry of spacetime.
	
	\subsection{FRW Cosmology}\label{cosmology}
We assume that our Universe is homogeneous and isotropic
	(according to the cosmological principle \footnote{It is noteworthy that the cosmological principle is a working assumption to provide a computable cosmology model and it is not rooted in a fundamental symmetry in physics. This means that by increasing the accuracy of observations, anomalies may be found that threaten the validity of this principle in some scales. A detailed discussion has been done in the review paper \cite{Aluri:2022hzs}.}) so that background
	geometry is described by the Friedmann-Robertson-Walker metric
	(FRW). Although is not excluded that in presence of the SLSB the
	bumblebee field $B_\mu$ may acquire a nonvanishing spatial
	orientation which, due to the breaking of the rotation symmetry, is a
	threat to the isotropy assumption of the Universe, we do not
	consider such a case (see Ref. \cite{paramos}).
	Instead, we assume that the bumblebee field obeys the following
	ansatz \cite{paramos}
	\begin{equation} \label{eqBform}
		B_\mu = \left(B(t),  0, 0, 0\right)\,.
	\end{equation}
Equivalently, the time-like background bumblebee field (\ref{eqBform}) can be viewed as a gradient of a time-dependent scalar. In any case, we deal with a quantity embedded in the background, whether $B_\mu$ or a scalar field. Even though for Lorentz-violating there are multiple scenarios, in this paper, we are interested in cosmologically constraining it in the same manner that bumblebee gravity addresses it i.e., the presence of a vector field in the background of spacetime. Note that in case of setting ansatz $B_\mu=(0,\overrightarrow{B})$ (as used in Ref. \cite{maluf}), which disturbs the homogeneity and isotropy properties of FRW metric, there is no longer a such possibility to consider a scalar field.
This choice preserves  the cosmological principle i.e., homogeneity and isotropy of the Universe, which evolves according to the (flat) FRW metric \footnote{Note that in case of taking a space-like bumblebee field ansatz, the FRW metric is no longer suitable and should be employed the Bianchini I metric, just like what was done in \cite{maluf}.}, described by the line element
	\beq\label{FRW}
	ds^2 = - dt^2 + a(t)^2 \left[ dr^2 + r^2   d\th^2 + r^2 \sin^2(\th)   d\ph^2 \right]\,,
	\eeq
	where $a(t)$ is the scale factor.
	From Eq. (\ref{eqBform}) it follows $B_{\mu\nu} = 0$, while the only nontrivial component of the bumblebee is (see Eq. (\ref{eqbumblebee}))
	\beq\label{eqVprime}
	\left[
	V' - {3(2 \chi + \xi) \over 2\ka} {\ddot{a}\over a} - {3\chi \over \ka}H^2
	\right] B = 0\,.
	\eeq
	For $B\neq 0$ one gets a relation between the dynamics of the potential and the scale factor.  Using (\ref{eqBform}), one gets the $00$ component of  (\ref{EqEinstein}),
	\beq \label{EqEinst00}
	H^2 [1- ( \xi + \chi ) B^2] = {1 \over 3} \kappa (\rho + V ) + ( \xi + 2 \chi ) H B \dot B\,,
	\eeq
	while the diagonal $ii$ components read
	\begin{eqnarray}\label{EqEinstij}
		&& \left( H^2  + 2 \frac{\ddot{a}}{a} \right)  [1- ( \chi + \xi ) B^2] = -\kappa p + \\ && + \kappa V + 4( \xi + \chi ) H B{\dot B}  + ( \xi + 2\chi ) ( {\dot B}^2 + B {\ddot B} )\,, \nonumber
	\end{eqnarray}
	where $H \equiv \dot{a}/a$ is the Hubble parameter. As one can see, the additional coupling cannot be absorbed in a redefinition of the parameters due to the presence of the two factors $\xi + \chi$ and $\xi + 2\chi$.
	Using the Bianchi identities, $\nabla_\mu (T^{\mu}_{\,\,\nu}+\frac{1}{\kappa}T^{(B)\mu}_{\,\,\,\,\nu})=0$, one obtains
	\begin{eqnarray}\label{EqContTot}
		\dot{\rho} + 3H(\rho+p) &=&  -\frac{1}{\kappa}\left[{\dot \rho}_B +3H(\rho_B+p_B)\right] \\
		& \equiv & -\frac{1}{\kappa}\Gamma_B\,, \nonumber
	\end{eqnarray}
	showing that there is an energy exchange between matter and the bumblebee field. In other words, $\Gamma_B$ refers to the amount of energy non-conservation in which its origin comes from the bumblebee background vector field.
	In (\ref{EqContTot}), $\rho_B$ and $p_B$ are defined in (\ref{TB00}) and (\ref{TBij}), respectively.
	%
	By using the equation of state for matter
	\begin{equation}
		p=w \rho\,,
	\end{equation}
	the general solution of (\ref{EqContTot}) is given by
	\begin{equation}
		\rho=\frac{\tilde \rho_0}{a^{3(1+w)}}+\delta_B
		\equiv \rho_0 + \delta_B\,,
	\end{equation}
	where $\rho_0={\tilde \rho}_0/a^{3(1+w)}$ is the standard energy density of matter in GR (${\tilde \rho_0}$ is a constant of integration), while $\delta_B$ accounts for bumblebee $B$-corrections
	\begin{equation}\label{deltaxi}
		\delta_B \equiv -\frac{1}{\kappa \, a^{3(1+w)}} \int \Gamma_B(t) a^{3(1+w)}(t) dt\,.
	\end{equation}
	From Eq. (\ref{EqEinst00}) we solve with respect to the potential $\kappa V$,
	\begin{eqnarray}\label{EqV}
		\kappa V &=& 3\left(H^2-\frac{\kappa\rho_0}{3}\right) -\kappa \delta_B - \\
		& & -3 H\,  B^2 \left[(\xi+\chi)H+(\xi+2\chi)\frac{\dot B}{B}\right]\,. \nonumber
	\end{eqnarray}
	Inserting (\ref{EqV}) into (\ref{EqEinstij}), one gets
	\begin{eqnarray}\label{19}
		-2H^2+2\frac{\ddot a}{a} &=& -\kappa (1+w)\rho_0-\kappa (1+w)\delta_B + \label{EqIntDiff} \\
		&+ &  (\xi+\chi)B^2\left[-2H^2+2\frac{\ddot a}{a}+4H\frac{\dot B}{B}\right] + \nonumber \\
		&+& (\xi+2\chi)B^2 \left[-3H\frac{\dot B}{B}+4\left(\frac{{\dot B}^2}{B^2}+\frac{\ddot B}{B}\right)\right]\,. \nonumber
	\end{eqnarray}
	This equation is an integro-differential equation. To find a solution, we make the following ansatz:
	\begin{equation}\label{ansatz}
		a(t)=a_0 t^\alpha\,, \quad
		B(t)=b \left(\frac{t}{\tilde t}\right)^\beta= b
		({\tilde M}\, t)^{\beta}\,.
	\end{equation}
	Here ${\tilde t}={\tilde M}^{-1}$ is some time/mass scale at which the bumblebee terms are effective and usually it is fixed around the Planck scale. Before proceeding with the calculation, it is helpful that we comment, due to the dependency of the output of our analysis on the ansatz (\ref{ansatz}), on the time evolution of the scale factor and bumblebee field.  The former comes from our interest in finding imprints of Lorentz-violating bumblebee vector field in the early Universe, particularly in BBN and Baryogenesis eras, in which it is expected the evolution of scale factor is of the power-law form, similar to the radiation-dominated epoch. Concerning the time evolution of the bumblebee vector field, one can show that, in essence, it is dependent on the form of the bumblebee potential $V \left( B^\mu B_\mu \pm b^2 \right)$. It is commonly proportional to $\left( B^\mu B_\mu \pm b^2 \right)^n$ which, with the derivative of its argument, has $V'\propto \left( B^\mu B_\mu \pm b^2 \right)^{n-1}$ \cite{paramos}. Besides, by putting the ansatz of scale factor (\ref{ansatz}) into Eq. (\ref{eqVprime}), we have $V'\propto t^{-2}$. Now it is clear that $B(t)\propto t^{\frac{-1}{n-1}}$. Re-expressing it in the form of $B(t)$ in (\ref{ansatz}), one obtains that the origin of the exponent $\beta$ indeed comes from the form of bumblebee potential. By passing  the case $n=1$ (linear form of bumblebee potential) we have $\beta>0$ and $<0$, if $n<1$ and $>1$, respectively. In this way, observational restriction derived in the next Sections on the exponent $\beta$, allows to rule out some of the power-law forms of bumblebee potential $\left( B^\mu B_\mu \pm b^2 \right)^n$.	
	
\begin{figure}[htp]
		\centering
		\includegraphics[scale=0.55]{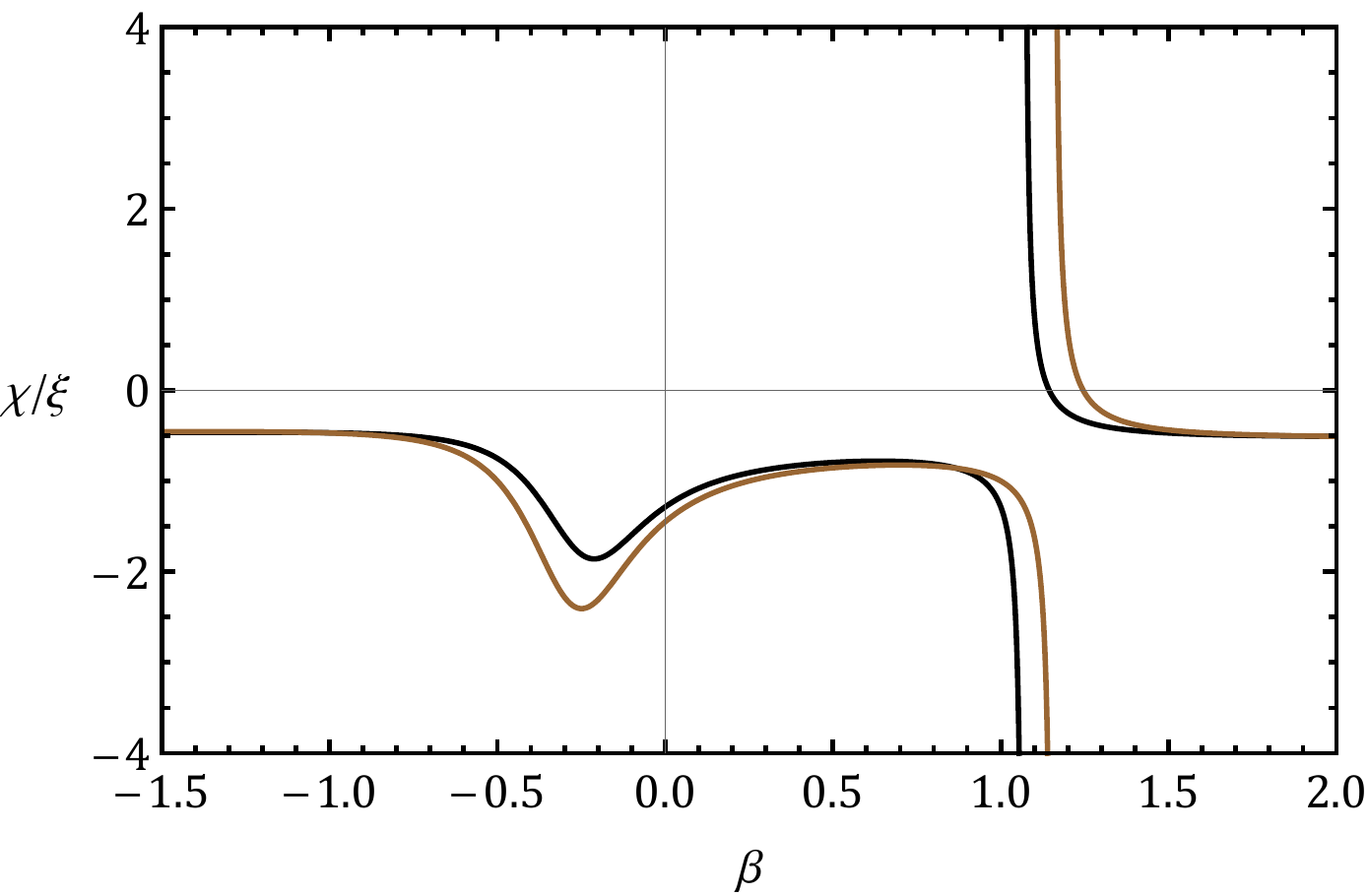}
		\caption{The behavior of dimensionless ratio $\chi/\xi$ in terms of exponent $\beta$ for epochs: RD (black curve) and MD (brown curve).} 
		\label{FigChiXiBeta}
	\end{figure}

	Plugging (\ref{ansatz}) into (\ref{deltaxi}) one infers
	\begin{equation}
		\label{deltaB}
		\delta_B = -\frac{b^2}{\kappa}\frac{3\alpha (B_B-A_B)}{3\alpha (1+w)+2\beta -2}\frac{{\tilde M}^{2\beta}}{t^{2-2\beta}}\,,
	\end{equation}
	where
	\begin{subequations}
		\begin{align}\label{ABdef}
			A_B \equiv& (\xi+\chi) (5\alpha^2+4\alpha\beta-\alpha)   \\ & + (\xi+2\chi)
			(3\alpha\beta+2\beta^2-\beta), \nonumber \displaybreak[0]\\[1ex]
			B_B \equiv& (\xi+\chi) (2\alpha\beta-2\alpha) \label{BBdef} \\ & +(\xi+2\chi) 
			(\alpha-1+2\beta(\beta-1))+2\chi\alpha \,. \nonumber
		\end{align}
	\end{subequations}
	In deriving (\ref{deltaB}) we used the relation
	\begin{eqnarray}\label{dotV}
		\kappa {\dot V} &=& \kappa V' \frac{d}{dt}(B_\alpha B^\alpha \pm b^2) \\
		&=& b^2 (3\alpha\beta) [(\xi+2\chi)(\alpha-1)+2\chi \alpha^2]\frac{{\tilde M}^{2\beta}}{t^{3-2\beta}}\,. \nonumber
	\end{eqnarray}
	Moreover, Eqs. (\ref{ansatz}) and (\ref{deltaB}) allow to rewrite (\ref{EqIntDiff}) in the form
	\begin{eqnarray}
		-\frac{2\alpha}{t^2} &=& -\kappa \frac{(1+w){\tilde \rho_0}}{a_0^{3(1+w)}t^{3\alpha (1+w)}}+ b^2 \, {\cal C}_{w, \alpha} \, \frac{{\tilde M}^{2\beta}}{t^{2-2\beta}}~, \label{EqEinstijnt}\\
		{\cal C}_{w, \alpha} & \equiv &  \frac{(1+w)3\alpha(B_B- A_B)}{3\alpha (1+w)+2\beta -2}+(\xi+\chi)(4 \alpha\beta -2\alpha)+  \nonumber \\
		& & + (\xi+2\chi) (8\beta^2-3\alpha \beta- 4\beta) \,.
		\label{CtildeB}
	\end{eqnarray}
	At the first glance, it can be seen that Eq. (\ref{EqEinstijnt}) meets its standard counterpart, if
	\begin{equation}\label{CB=0}
		{\cal C}_{w, \alpha} = 0\,.
	\end{equation}
	By setting the values of the adiabatic index (equation of state parameter) $w$ and the exponent of the scale factor $\alpha$ from the standard cosmology, thereby, one can interpret Eq. (\ref{EqEinstijnt}) as an equation for determining dimensionless ratio $\chi/\xi$ in terms of $\beta$. The values $\{w=1/3, \alpha=1/2\}$ and $\{w=0, \alpha=2/3\}$ correspond, respectively, to {\it Radiation Dominated (RD) era} and {\it Matter Dominated (MD) era}, and Eqs. (\ref{EqEinstijnt}), (\ref{CtildeB}), and (\ref{CB=0}) give
	\begin{equation}\label{R}
		\frac{\chi}{\xi}=\frac{9+18\beta +14\beta^2-32\beta^3}{-7-28\beta-36\beta^2+64\beta^3}~,
	\end{equation}
	and
	\begin{equation}
		\frac{\chi}{\xi}=\frac{29+51\beta +30\beta^2-72\beta^3}{-20-78\beta-42\beta^2+72\beta^3}~.
	\end{equation} 
	Using the above relations $\chi/\xi$ vs $\beta$ is reported in Fig. \ref{FigChiXiBeta}. 
	As we can see, the exponent $\beta$ may assume all values (both positive and negative) except for some ones around $\beta=1$ in which ratio $\chi/\xi$ diverges. These results will be used in the next Sections when we will discuss BBN and the gravitational baryogenesis.
	
	Concerning the {\it Cosmological constant (current era)} also one easily realizes that by setting $w=-1$, then a solution of (\ref{EqEinstijnt}) admits as solution $B(t)= constant$ \cite{paramos}. Let us emphasize that, due to our interest in BBN and Baryogenesis epochs, throughout this paper, the last two cases are out of our attention, and the main concentration is the first case i.e., Eqs. (\ref{R}) for the RD era.
	
	At the end of this Section, we report the trace of the energy-momentum
	tensor of the bumblebee field in the FRW universe as follows
	\begin{eqnarray}\label{TraceTBFRW}
		T^{(B)} &=& 4\kappa V+ \label{TraceTBgen} \\
		&+& 3 B^2\Big[ (\xi+\chi)\left(2H^2+\frac{\ddot a}{a}+4H\frac{\dot B}{B}\right)+\nonumber \\
		&+& (\xi+2\chi)\left(H\frac{\dot B}{B}+\frac{{\dot B}^2}{B^2}+\frac{\ddot B}{B}\right)\Big] \nonumber \\
		&=& 4\kappa V + 3b^2 \Big[ (\xi+\chi) \alpha (3\alpha+4\beta-1)+  \label{TBFRW} \\
		& & + (\xi+2\chi)\beta (\alpha+2\beta-1)
		\Big] \frac{{\tilde M}^{2\beta}}{t^{2-2\beta}}\,. \nonumber
	\end{eqnarray}
	The components $00$ and $ii$ of the energy-momentum tensor above address energy density and pressure arising from the presence of bumblebee vector fields in the background (see (\ref{TB00}) to (\ref{pBFRW})). As the final word, in Appendix (\ref{C}), ansatz (\ref{ansatz}) is supported by linear dynamic system analysis.
	
	\section{BBN  in bumblebee cosmology}
	\label{BBN}
	
	In the Section, we examine the constraint on $\xi b^2$ coming from BBN. For our aim, the analysis here discussed to infer such a bound is enough.
	
\begin{figure*}[ht!]
		\begin{tabular}{c}
			\includegraphics[scale=0.42]{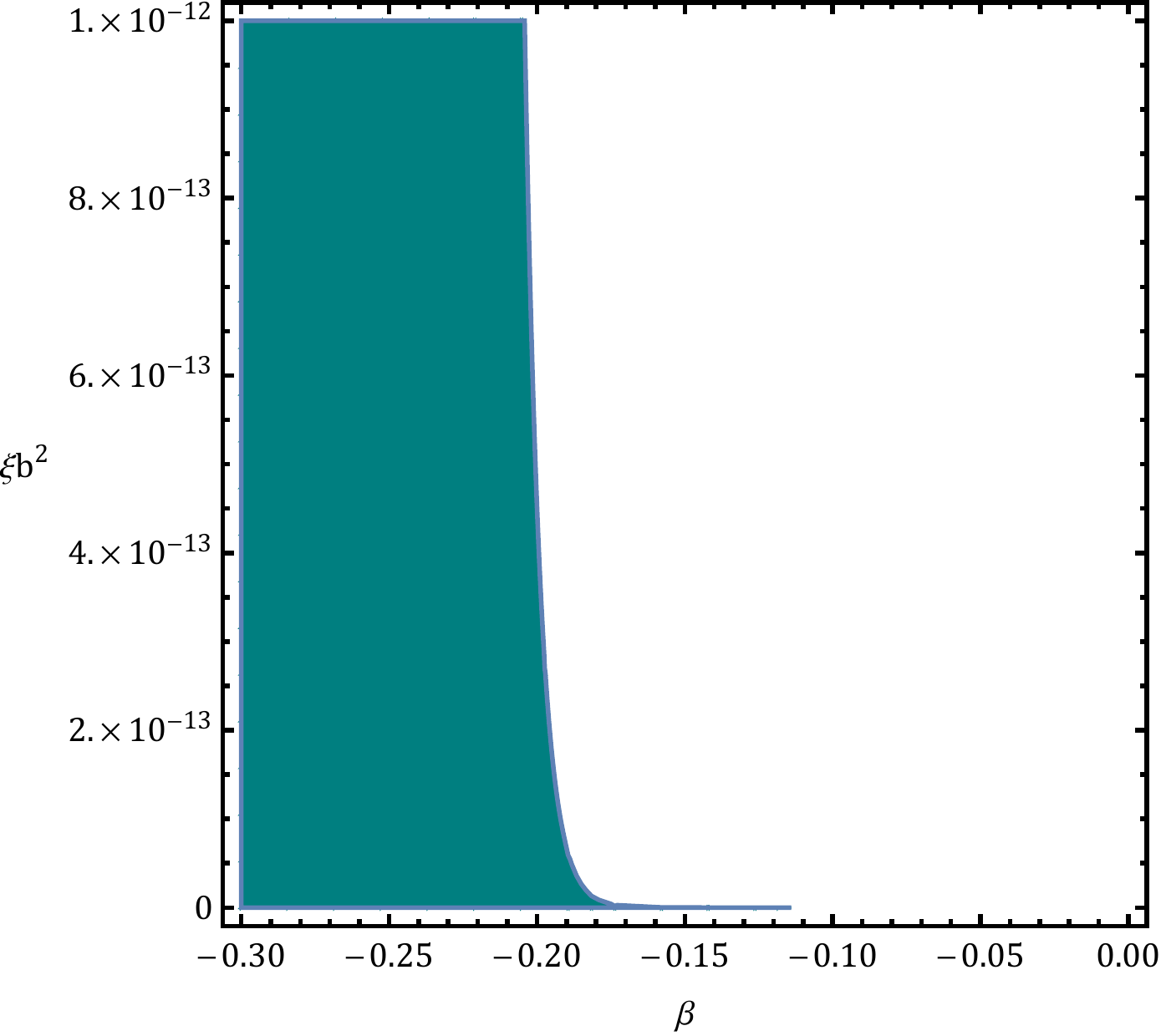}
			\includegraphics[scale=0.42]{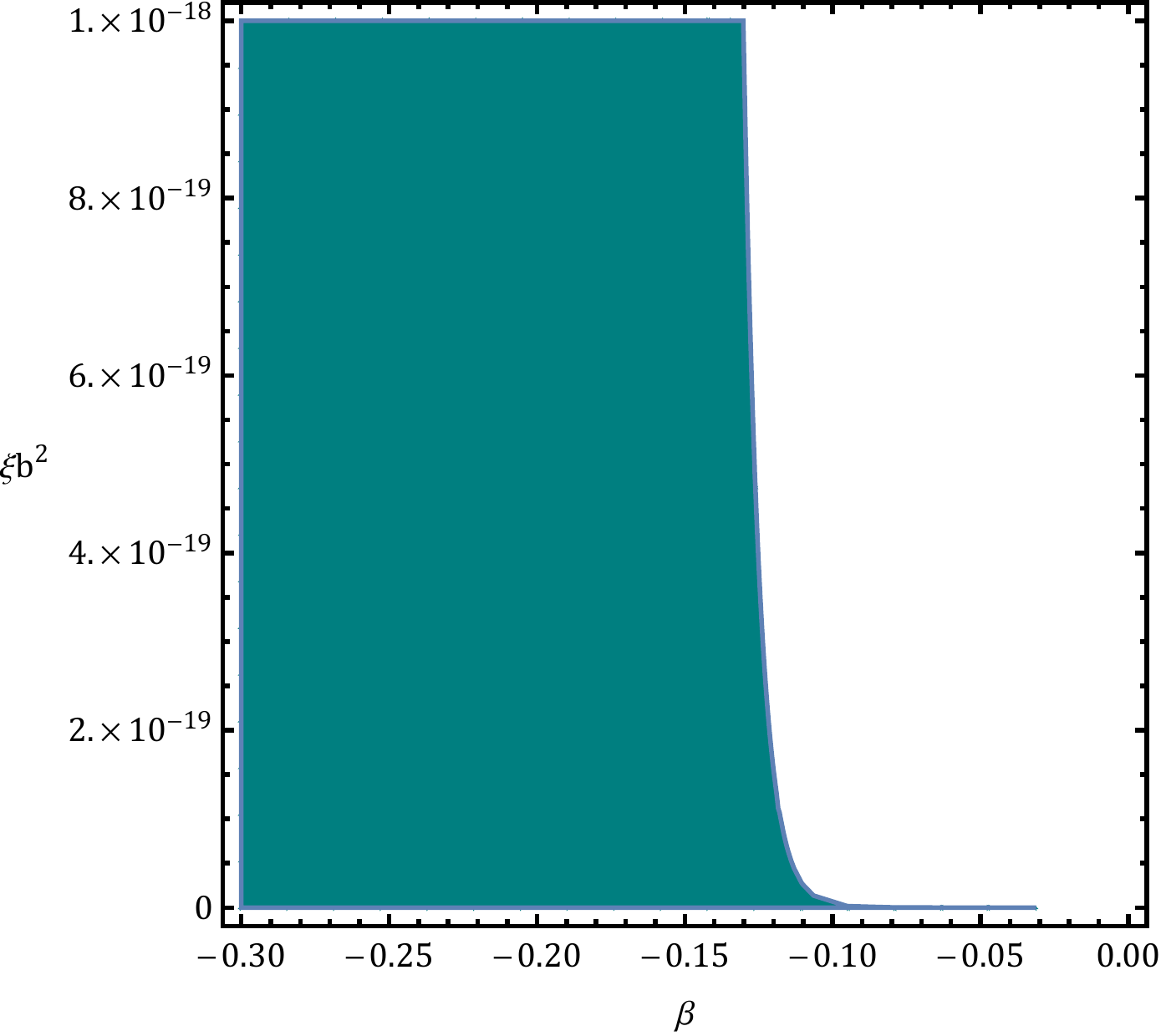}
			\includegraphics[scale=0.42]{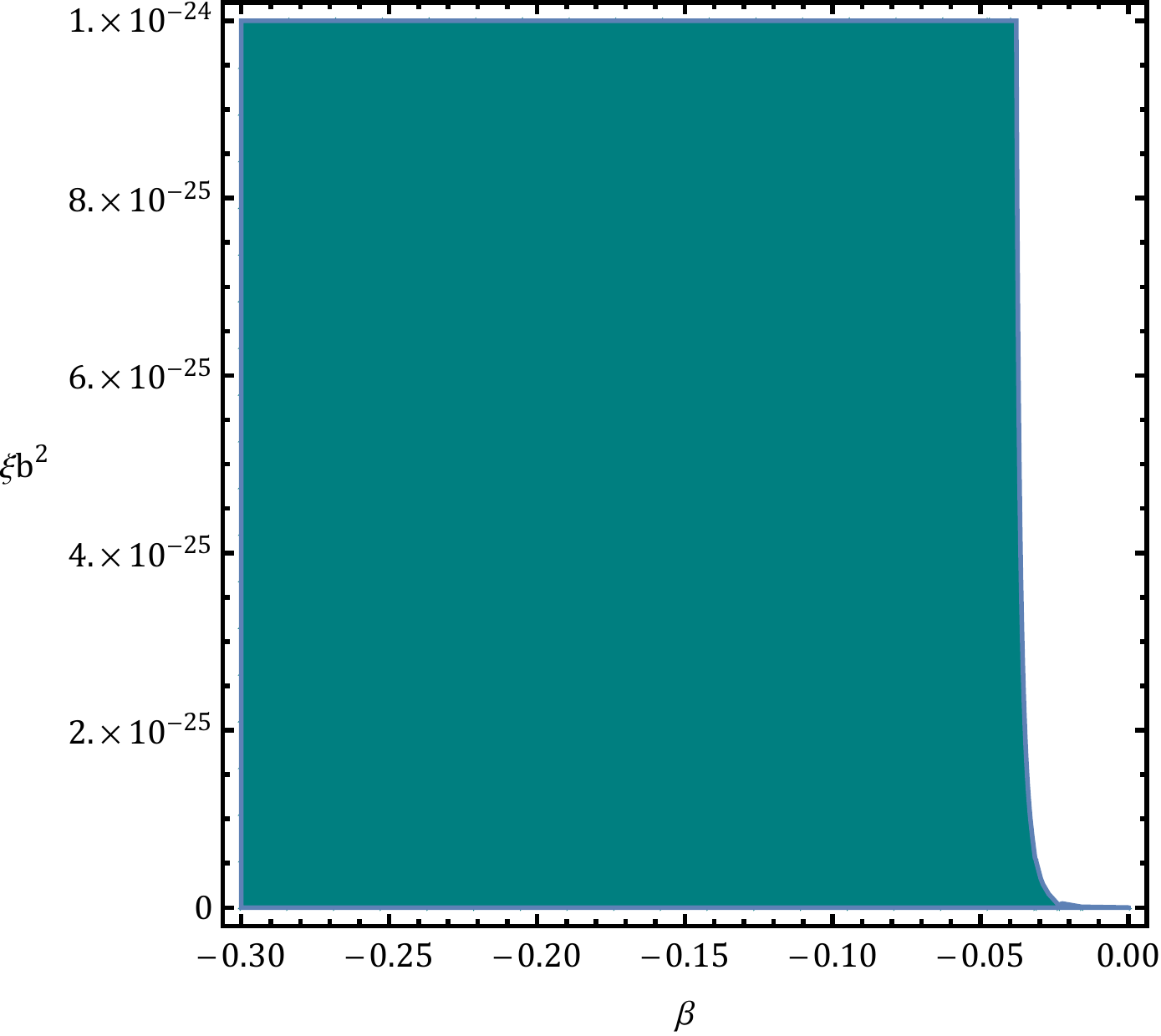}\\
			\includegraphics[scale=0.38]{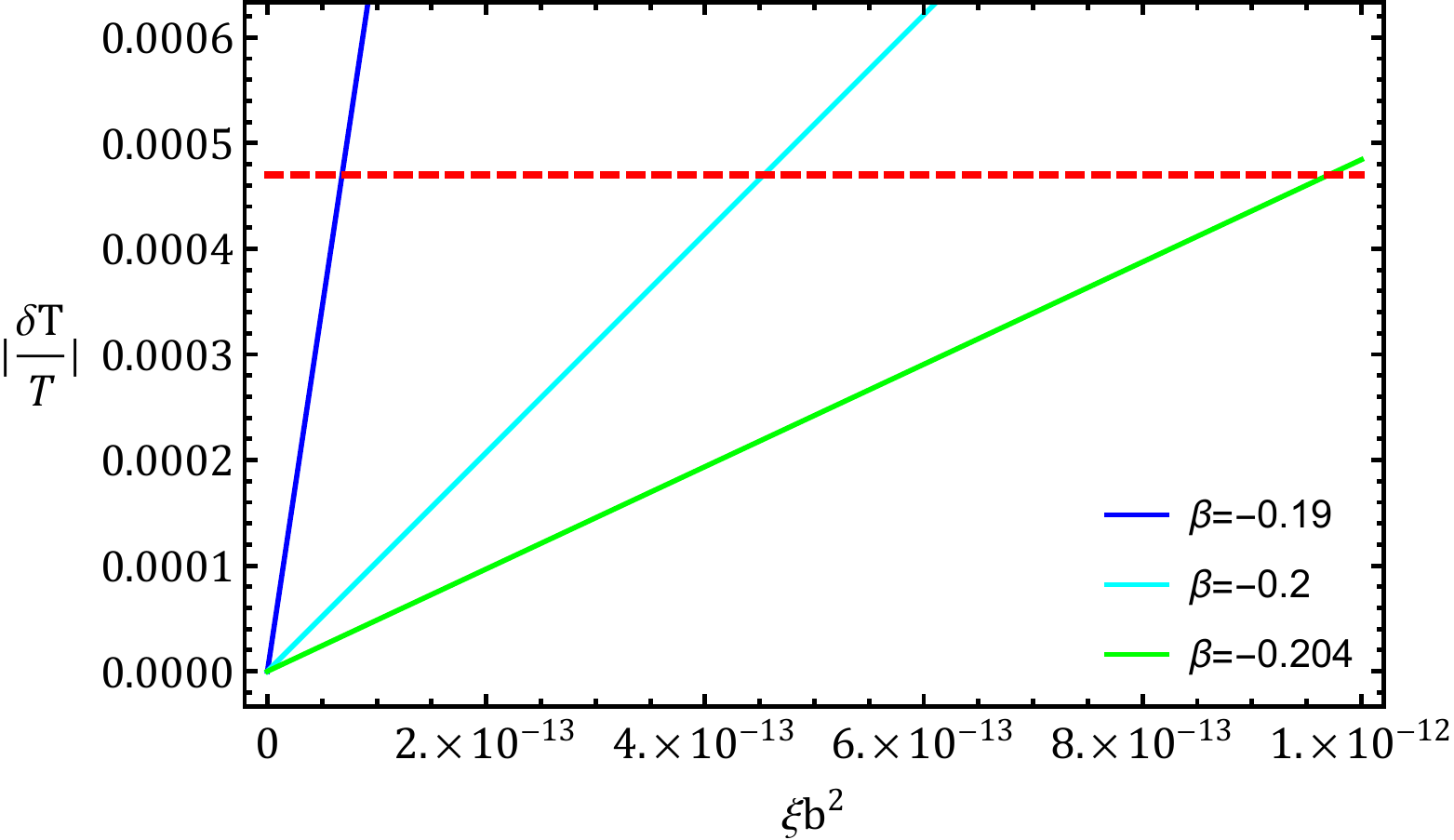}
			\includegraphics[scale=0.38]{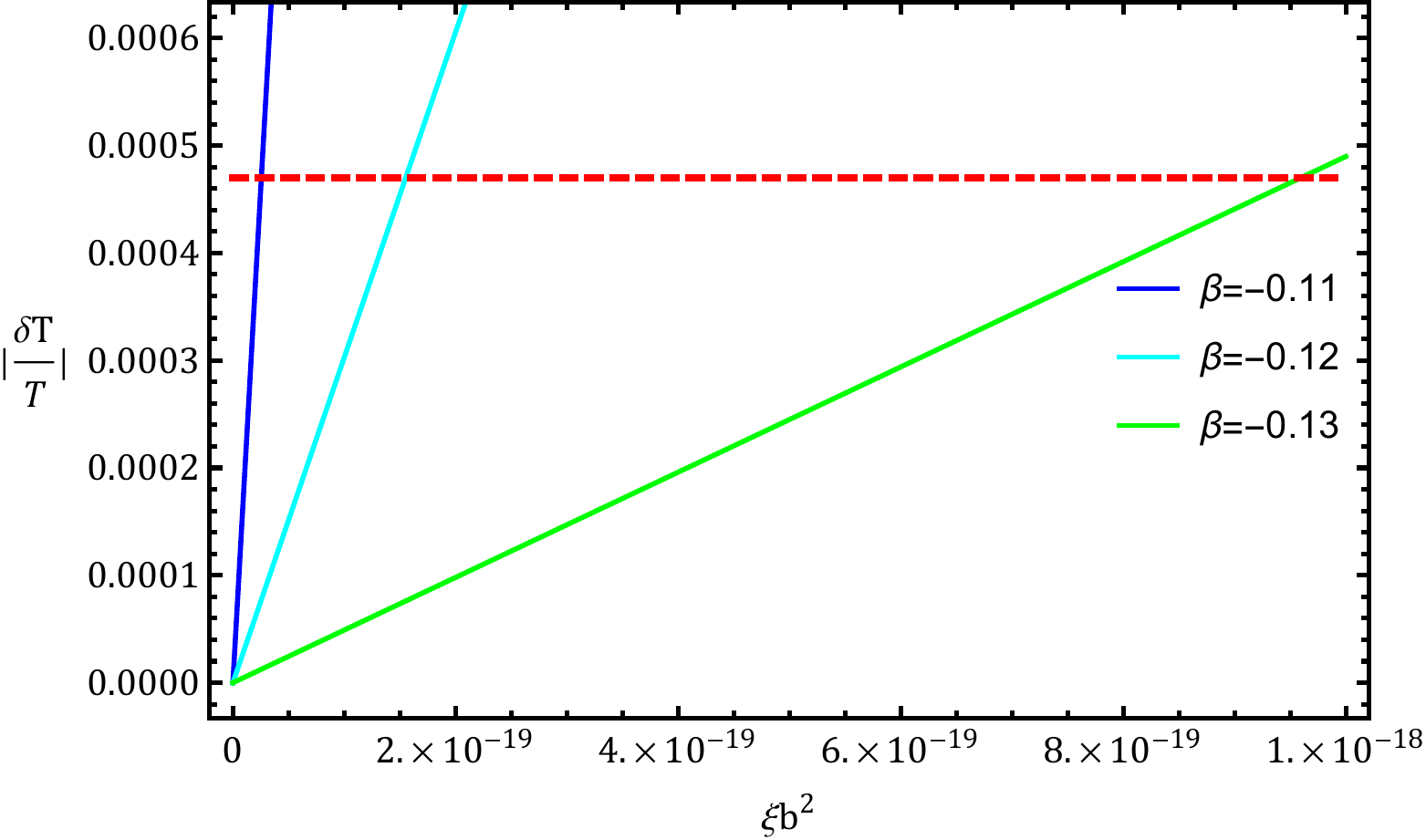}
			\includegraphics[scale=0.38]{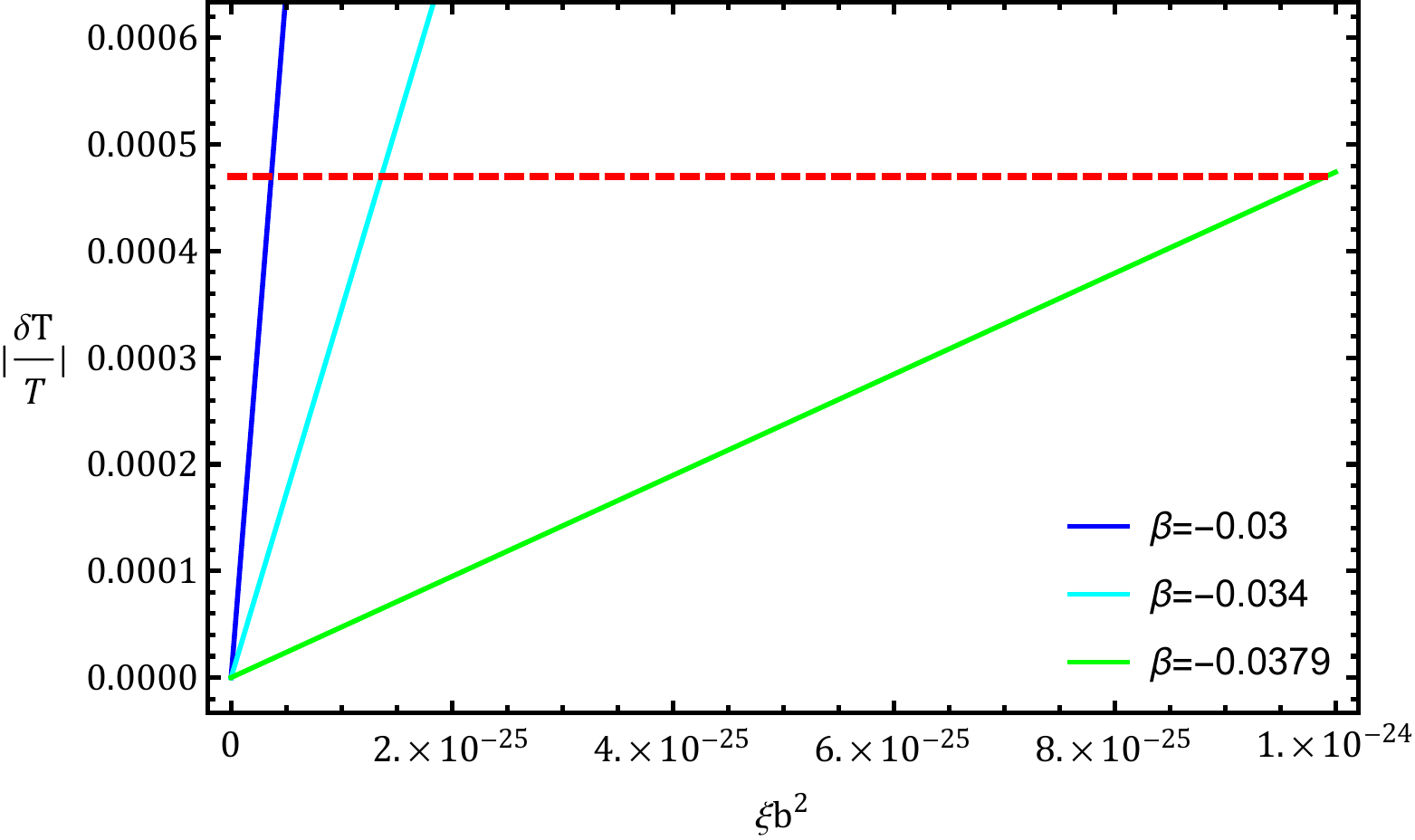}
		\end{tabular}
		\caption{Up row: Regions of existence in the $\beta-\xi b^2$ parameter space which satisfies the upper bound (\ref{deltaTTbound}). Bottom row: $\left|\frac{\delta {T}_f}{{T}_f}\right|$ from  (\ref{deltaTTboundG}) in terms of $\xi b^2$ for optional values of $\beta$ which put in correspond allowed region. Here we set numerical values: ${\tilde M}\sim 10^{19}$GeV$\sim M_P$, $T_{f}\sim 6\times10^{-4}$GeV, and $g_*=106.7$.}
		\label{NNB1}
	\end{figure*}

\begin{figure*}[ht!]
		\begin{tabular}{c}
			\includegraphics[scale=0.42]{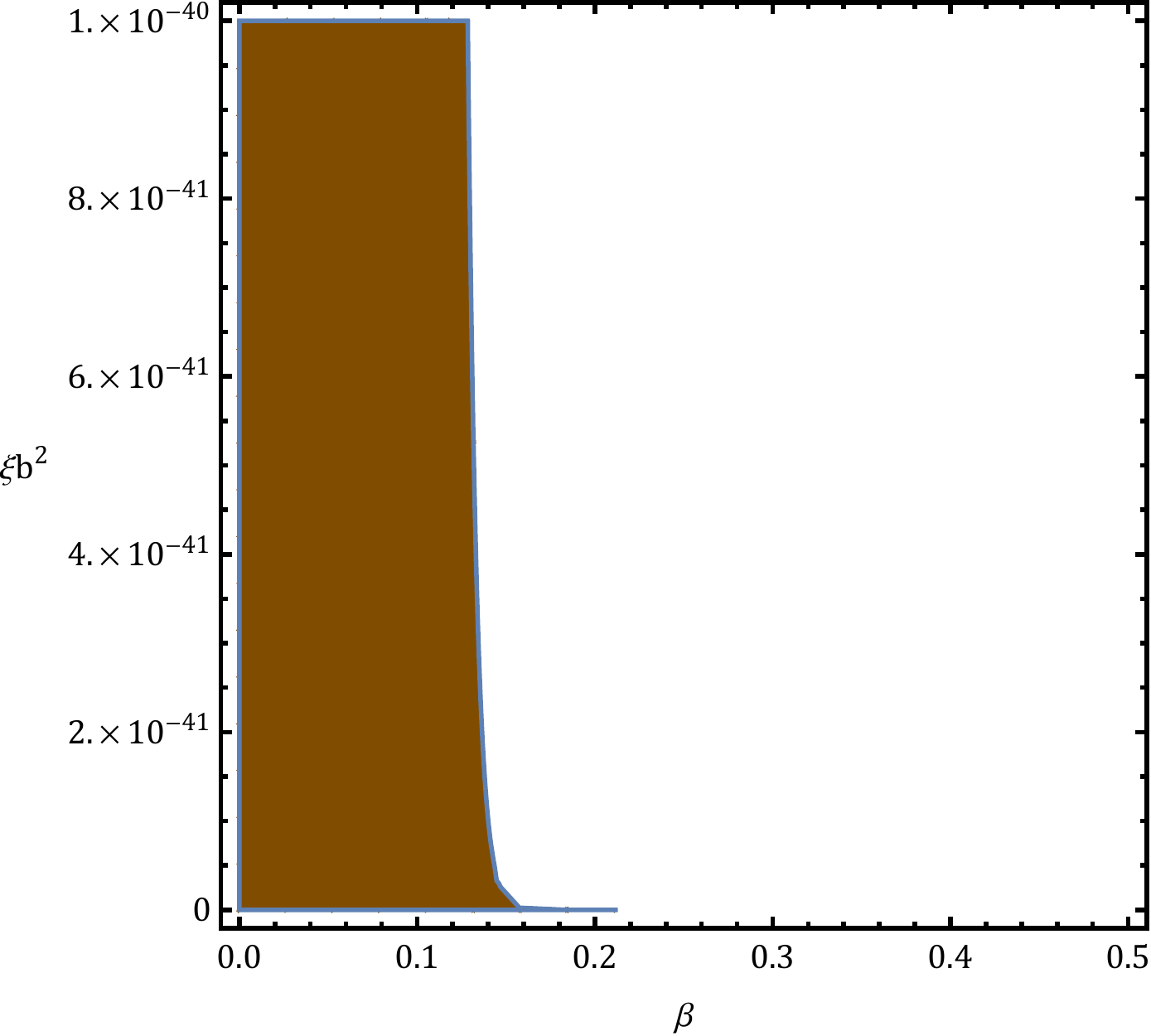}
			\includegraphics[scale=0.42]{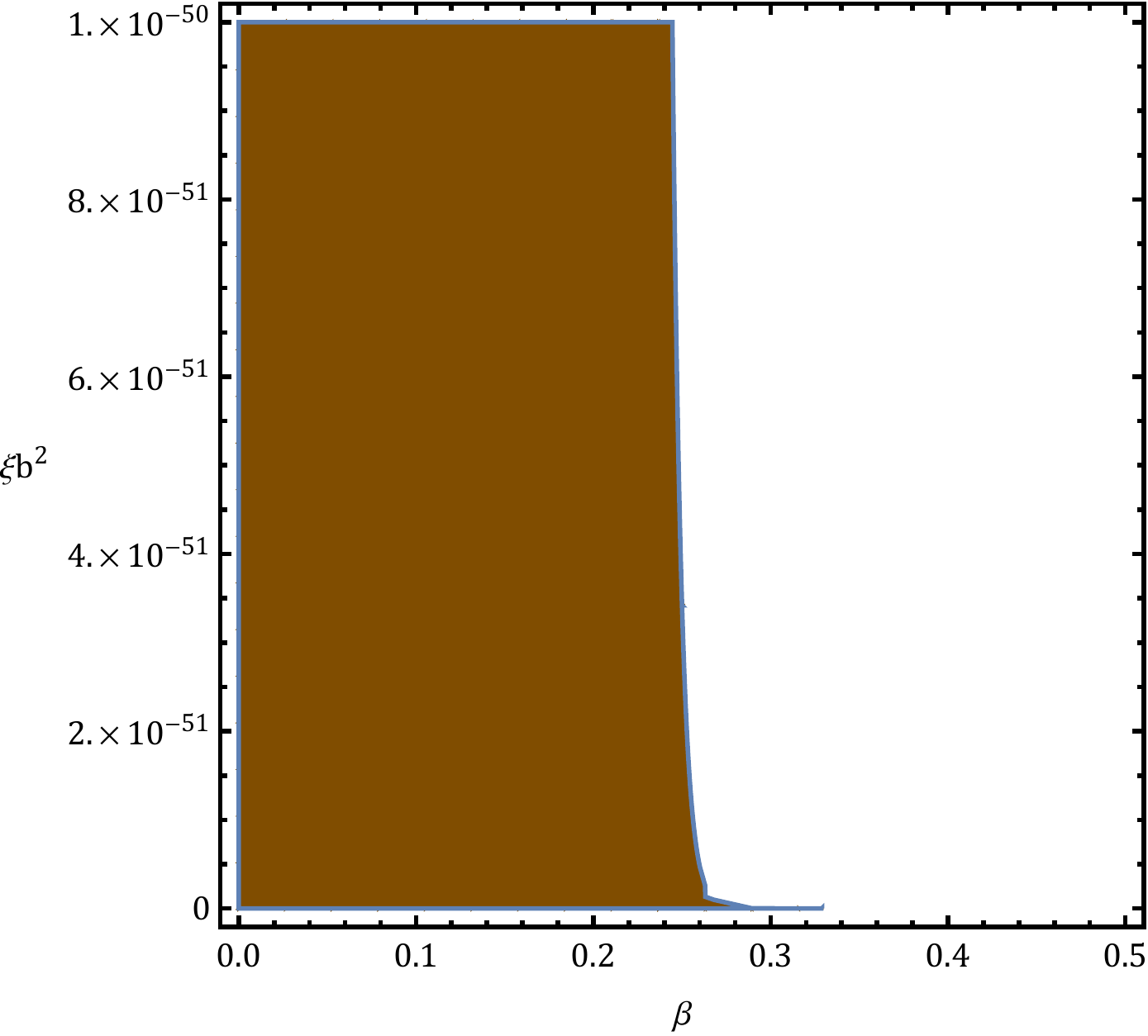}
			\includegraphics[scale=0.42]{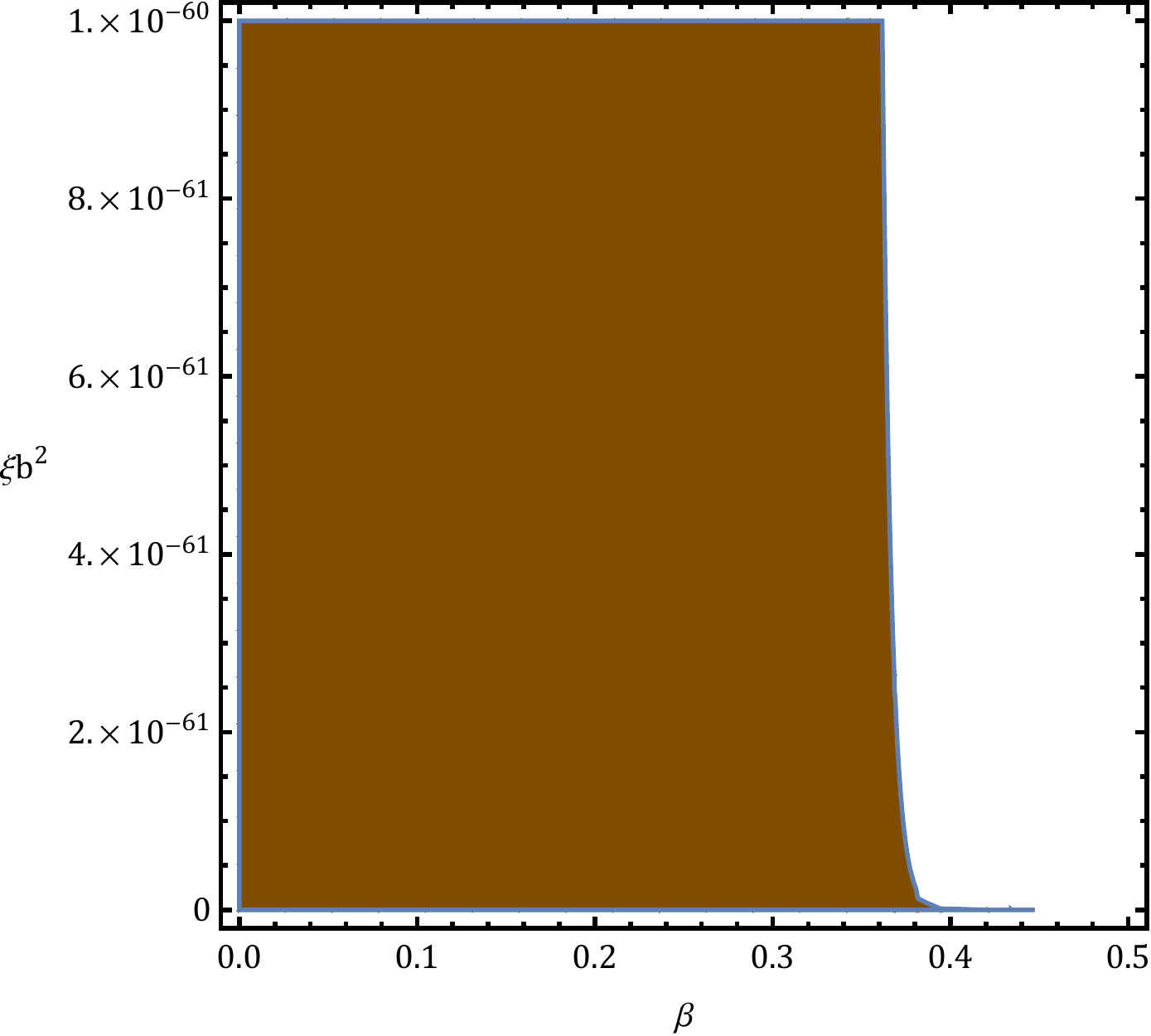}\\
			\includegraphics[scale=0.38]{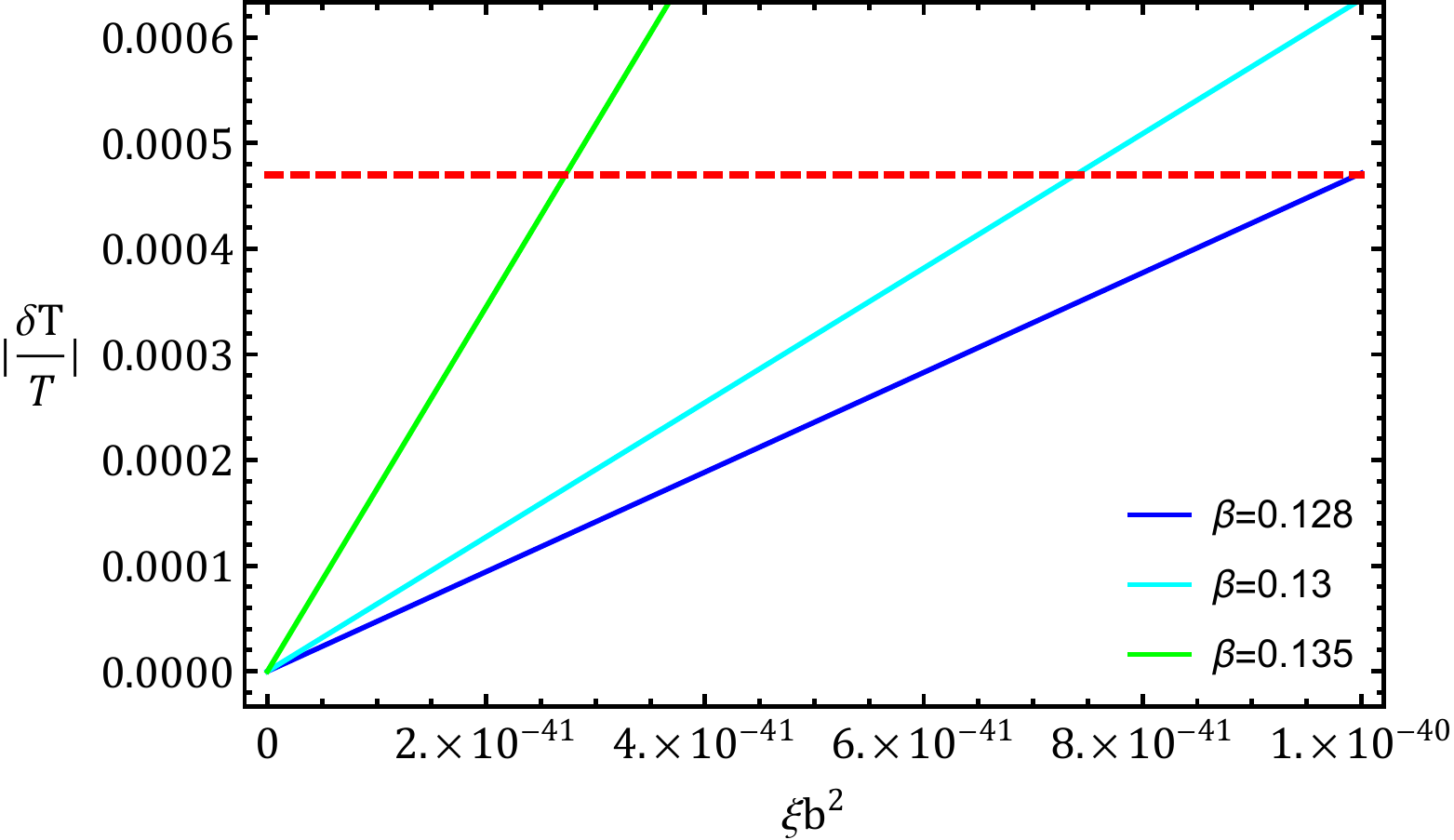}
			\includegraphics[scale=0.38]{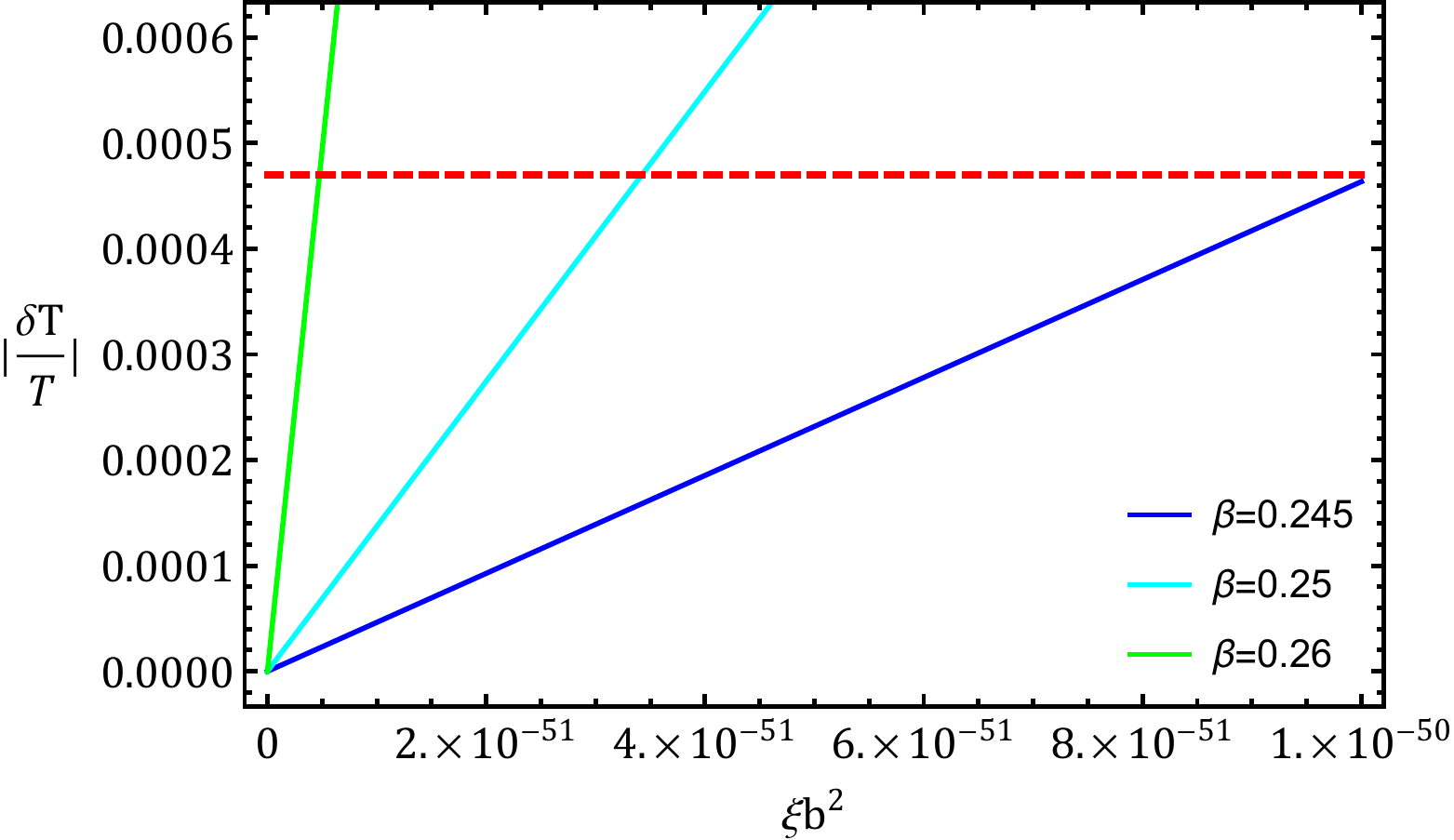}
			\includegraphics[scale=0.38]{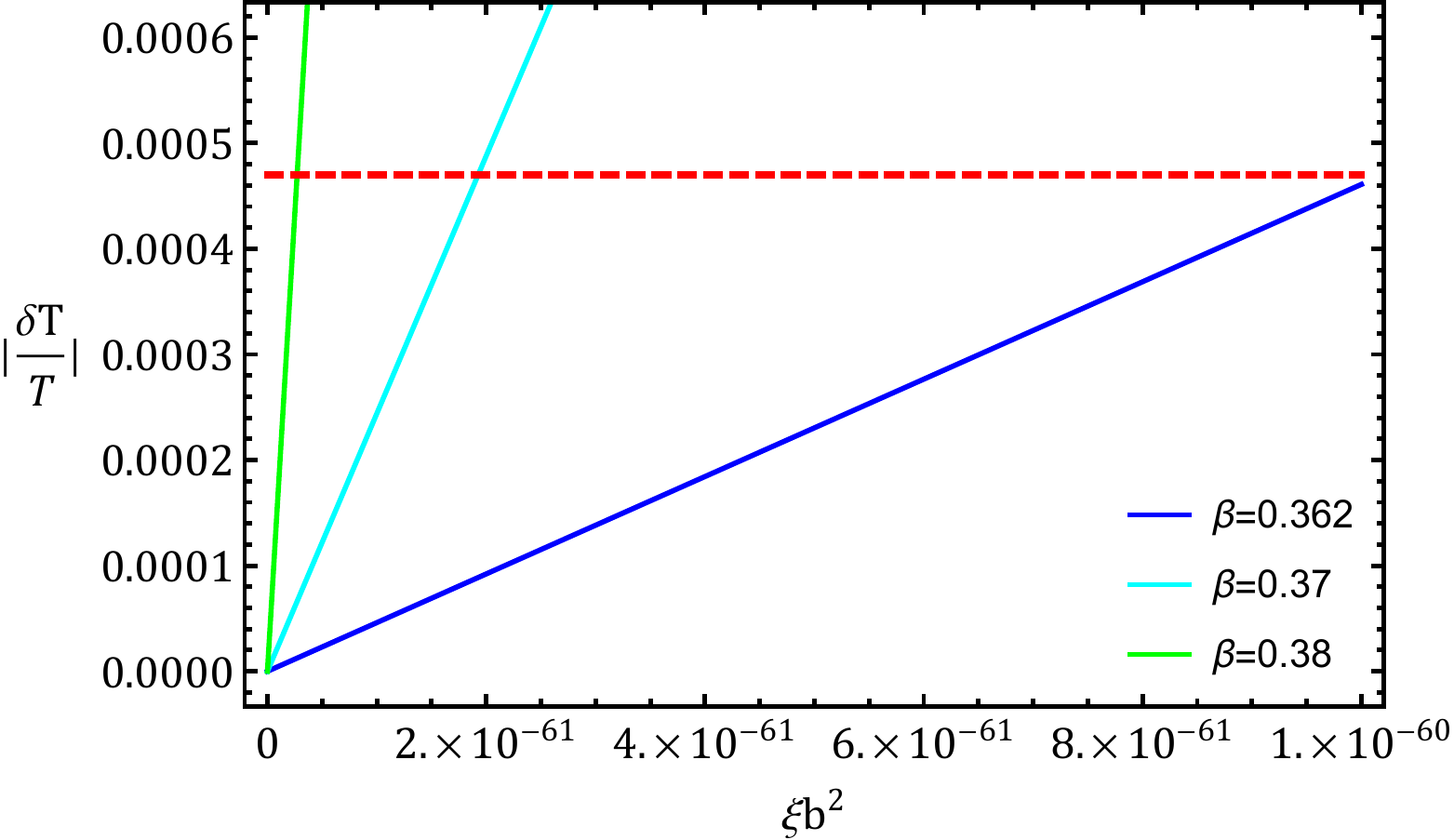}
		\end{tabular}
		\caption{Same as the Fig.~(\ref{NNB1}) but for $\beta>0$.}
		\label{NNB2}
	\end{figure*}

	BBN starts during the radiation
	dominated era \cite{kolb,bernstein,olivePDGroup2014}.
	The neutron abundance can be calculated via the
	conversion rate of protons into neutrons
	\[
	\lambda_{pn}({T})=\lambda_{n+\nu_e\to p+e^-}+\lambda_{n+e^+\to p+{\bar
			\nu}_e}+\lambda_{n\to p+e^- +
		{\bar \nu}_e}\,,
	\]
	and its inverse $\lambda_{np}({T})$, thus the total rate reads
	\begin{equation}\label{Lambda}
		\Lambda({T})=\lambda_{np}({T})+\lambda_{pn}({T})\,.
	\end{equation}
	From (\ref{Lambda}) one gets (see (\ref{LambdafinApp}))
	\begin{equation}\label{Lambdafin}
		\Lambda(T) =4 A\, {T}^3(4! {T}^2+2\times 3! {\cal Q}{T}+2!
		{\cal Q}^2)\,,
	\end{equation}
	where ${\cal Q}$ stands for the difference between neutron and proton mass,  ${\cal Q}=m_n-m_p$, while the numerical factor $A$ is given by $A=1.02 \times
	10^{-11}$GeV$^{-4}$. The ${}^4 He$ primordial mass
	fraction is estimated by using the relation \cite{kolb}
	\begin{equation}\label{Yp}
		Y_p\equiv \lambda \, \frac{2 x(t_f)}{1+x(t_f)}\,,
	\end{equation}
	in which $\lambda=e^{-(t_n-t_f)/\tau}$ ($t_f$   corresponds to the time of the freeze-out of the weak interactions, while $t_n$ to the time of the freeze-out of the nucleosynthesis),
	$\tau$ is the neutron mean lifetime defined in (\ref{rateproc3}), and, finally,
	$x(t_f)=e^{-{\cal
			Q}/{\cal T}(t_f)}$ is the neutron-to-proton equilibrium ratio.
	%
	%
	The variation of the freezing temperature ${T}_f$ induces a deviation from the fractional mass $Y_p$ given by
	\begin{equation}\label{deltaYp}
		\delta
		Y_p = Y_p\left[\left(1-\frac{Y_p}{2\lambda}\right)\ln\left(\frac{2\lambda}{Y_p}
		-1\right)-\frac{2t_f}{\tau}\right]
		\frac{\delta {T}_f}{{T}_f}\,,
	\end{equation}
	where $\delta {T}(t_n)=0$ has been used (it comes from the fact that ${T}_n$ is fixed by the deuterium
	binding energy
	\cite{torres,Lambiase3}).  Observations provide an estimation of $Y_p$ of baryon converted to ${}^4 He$ given by
	\cite{coc,altriBBN1,altriBBN2,altriBBN3,altriBBN4,altriBBN5,altriBBN6}
	\begin{equation}\label{Ypvalues}
		Y_p=0.2476\,, \qquad |\delta Y_p| < 10^{-4}\,.
	\end{equation}
	Combining Eqs. (\ref{Ypvalues}) and (\ref{deltaYp}) one gets
	\begin{equation}\label{deltaTTbound}
		\left|\frac{\delta {T}_f}{{T}_f}\right| < 4.7 \times 10^{-4}\,.
	\end{equation}
	For our aim, we rewrite the expansion rate of the BG-based
	universe at hand i.e., Eq. (\ref{EqEinst00}) in the form
	\begin{eqnarray}\label{H+H1}
		H&=&H_{GR} \sqrt{1+\frac{\rho_B}{\kappa \rho}} \equiv H_{GR}+\delta H\,,  \\
		\delta H&=&H_B \equiv  \left(\sqrt{1+\frac{\rh_B}{\kappa \rho}}-1\right)H_{GR}\,,
	\end{eqnarray}
	where $H_{GR}=\displaystyle{\sqrt{\kappa \rho}}$ is the expansion rate
	of the Universe in the standard cosmological model, ${\displaystyle \rho=\frac{\pi^2}{30}g_* {T}^4}$, and $\rho_B$ is defined in (\ref{TB00}). The relation $\Lambda= H$ gives the
	freeze-out temperature
	${T}={T}_f\left(1+\frac{\delta {T}_f}{{T}_f}\right)$, with ${T}_f\sim 0.6$ MeV obtained from $H(T_f)=\Lambda\simeq q {T_f}^5$, which $q\simeq 9.6\times 10^{-36}\, \text{GeV}^{-4}=\frac{9.6 \times 10^{40}}{M_P^4}$. Given that the deviation given raised of background bumblebee vector field from standard cosmology will lead to a deviation in the freeze-out temperature, thereby,  by taking $\delta H=\delta H(T_f)$ into account, we arrive at
	\begin{equation}\label{deltaTTboundG}
		\frac{\delta {T}_f}{{T}_f}  =
		\frac{\left(\sqrt{1+\frac{\rho_B}{\kappa \rho}}-1\right)H_{GR}}{5q {T}_f^5}
		\simeq \frac{\rho_B}{\kappa \rho}\frac{H_{GR}}{10 q
			{T}_f^5}\,,
	\end{equation}
	The last term in (\ref{deltaTTboundG}) follows from this reasonable demand which $\rho_B \ll \rho$.
	%
	%
	We then get
	\begin{equation}\label{rhoBoverkrho}
		\frac{\rho_B}{\kappa \rho}= \xi b^2 \Pi_{\xi, \chi} \left(\frac{45}{16\pi^3 g_*}\right)^\beta \left(\frac{{\tilde M} M_P}{T_f^2}\right)^{2\beta}\,,
	\end{equation}
	and
	\begin{equation}\label{HGRoverT}
		\frac{H_{GR}}{10 \,q\, T_f^5}=\frac{1}{10}\sqrt{\frac{8\pi^3g_*}{30}}\, \frac{1}{qM_P T_f^3} =
		\frac{10^{-40}}{48}\sqrt{\frac{\pi^3 g_*}{15}}\, \left(\frac{M_P}{T_f}\right)^3\,,
	\end{equation}
	where $\Pi_{\xi, \chi}$ is defined as
	\begin{eqnarray}\label{Pixichi}
		\Pi_{\xi, \chi} \equiv  - \frac{
			\left(1+\frac{\chi}{\xi}\right) (2\alpha\beta-2\alpha)+ \left(1+2\frac{\chi}{\xi}\right)(8\beta^2-3\alpha\beta-4\beta)}{3(1+w)}\,.
	\end{eqnarray}
	Note that here we should set the values of $\alpha$ and $\chi/\xi$
	from the RD epoch. By imposing the upper bound
	(\ref{deltaTTbound}) on Eq. (\ref{deltaTTboundG}), in Figs.
	\ref{NNB1} and \ref{NNB2} (up rows) we illustrate the parameter
	space plots in terms of $\beta-\xi b^2$ which address the allowed
	regions in which the upper bound (\ref{deltaTTbound}) satisfies.
	Also, in the bottom rows, we plot $\left|\frac{\delta
		{T}_f}{{T}_f}\right|$ in terms of $\xi b^2$ for optional values of
	$\beta$. As is evident, the upper bounds on $\xi b^2$ are
	sensitive to setting the value of exponent parameter $\beta$
	($B(t)\sim t^\beta$) so that for the negative case, the constraint on $\xi b^2$ is getting tighter as the value of $|\beta|$ gets smaller. For the case of $\beta>0$ also this statement works i.e.,
	increasing the value of $\beta$ results in the upper bound on $\xi
	b^2$ shifts to lower ones. Concerning the negative case, for
	example by setting $\beta=-0.204$ one obtains the constraint
	$\lesssim10^{-12}$ for $\xi b^2$, while for $\beta\approx-0.038$
	it falls to range $\lesssim10^{-24}$ which is $12$ order of
	magnitude more stringent than the former. As a result, it is
	expected that the corresponding upper bounds on $\xi b^2$ become
	even tighter than $10^{-24}$, if $\beta>-0.038$ (see the right
	panel in the bottom row of Fig. \ref{NNB1}). Concerning the case
	$\beta>0$ a comment is in order. The values of $\beta>0$ imply
	that the bumblebee field grows with the cosmic time $t$, or,
	equivalently, increases as the temperature decreases. Despite that,
	this scenario guides us to tight upper bounds for  $\xi b^2$
	(see Fig. \ref{NNB2}), they can not be reliable. In other words,
	these very tight constraints, in essence, come from the scenario
	that seems not cosmologically favourite since commonly one expects
	that the Lorentz violation terms are merely effective in the early
	universe, at high temperatures.
	
	
	\begin{figure*}[ht!]
		\begin{tabular}{c}
			\includegraphics[scale=0.45]{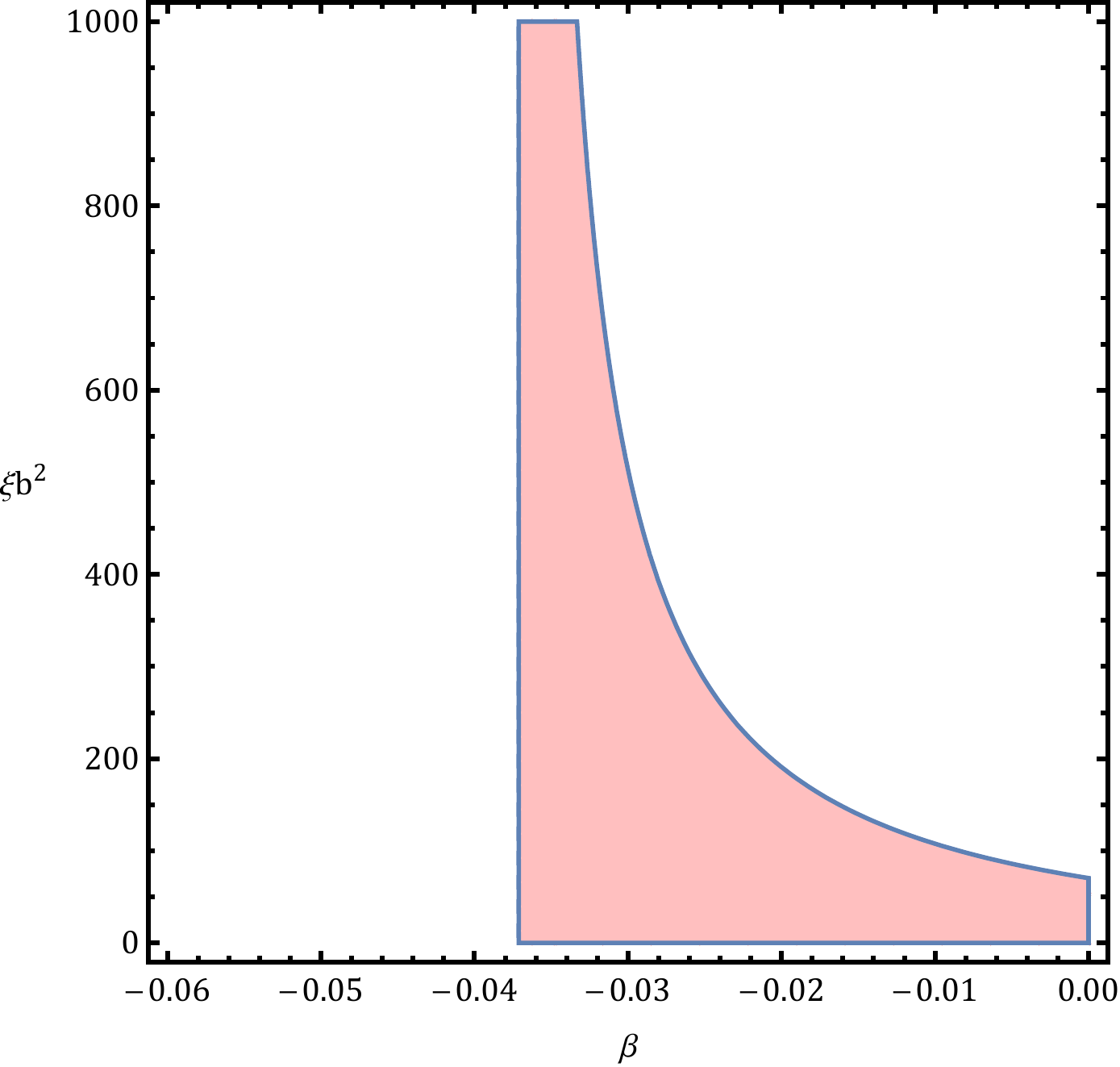}~~
			\includegraphics[scale=0.45]{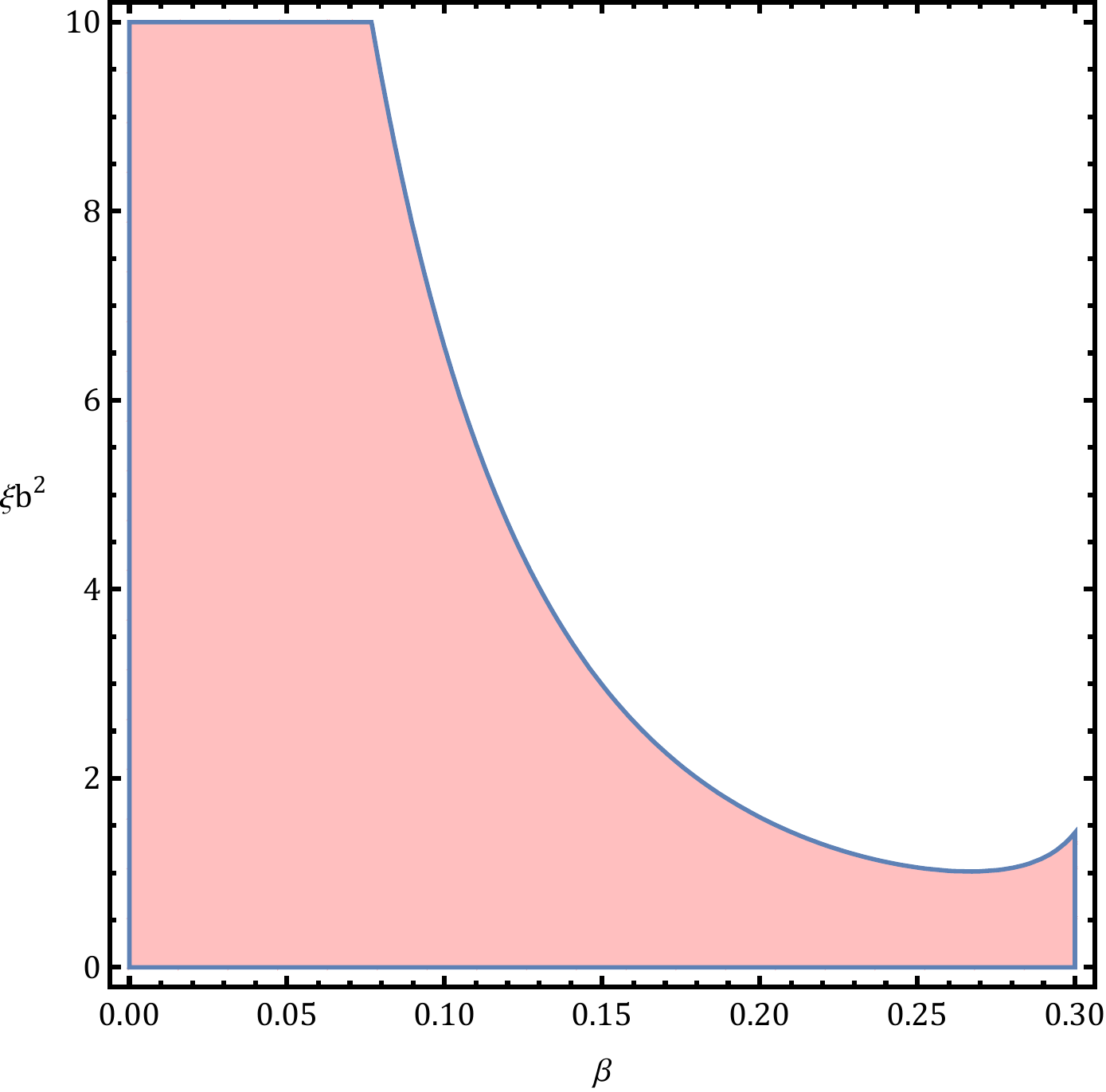}\\
			\includegraphics[scale=0.45]{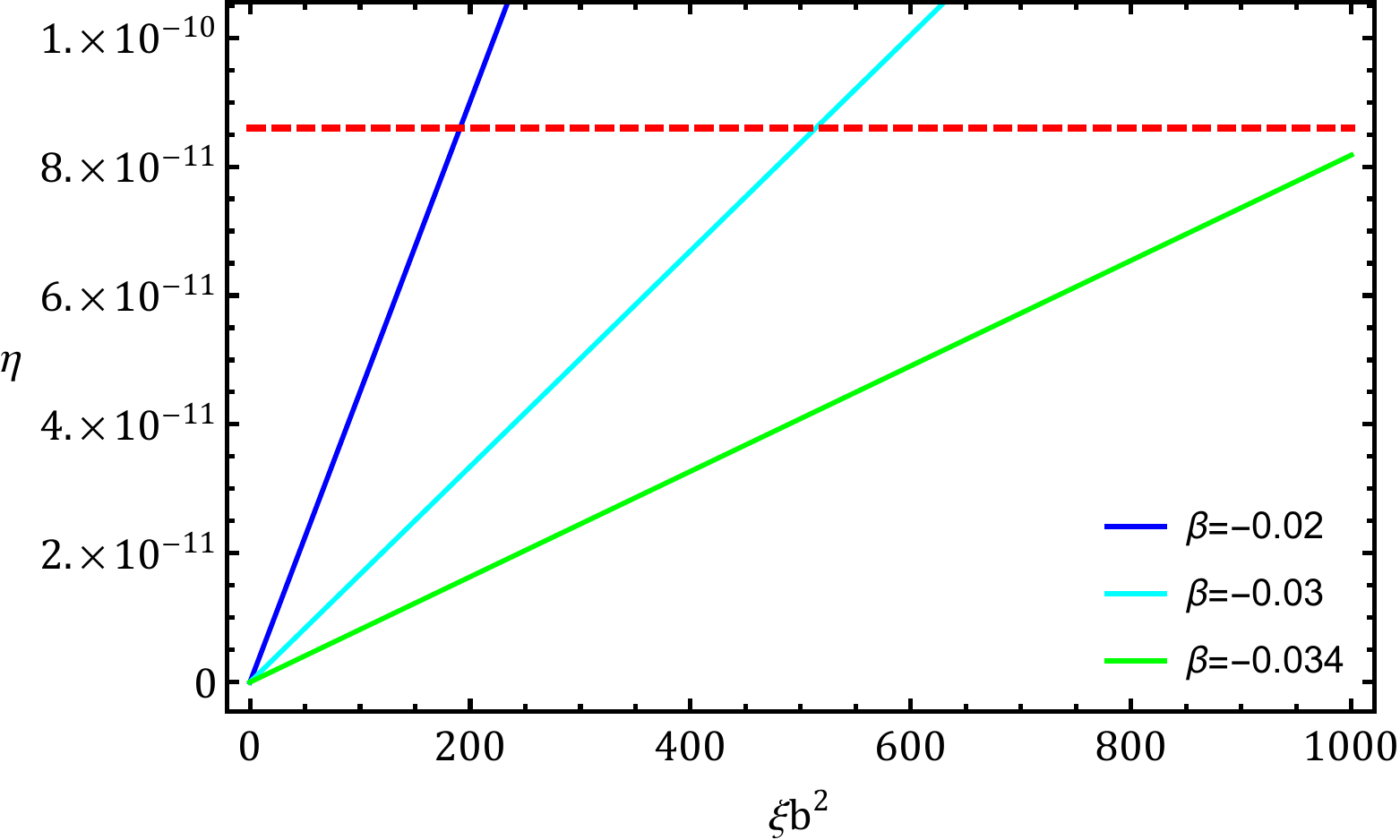}~~
			\includegraphics[scale=0.45]{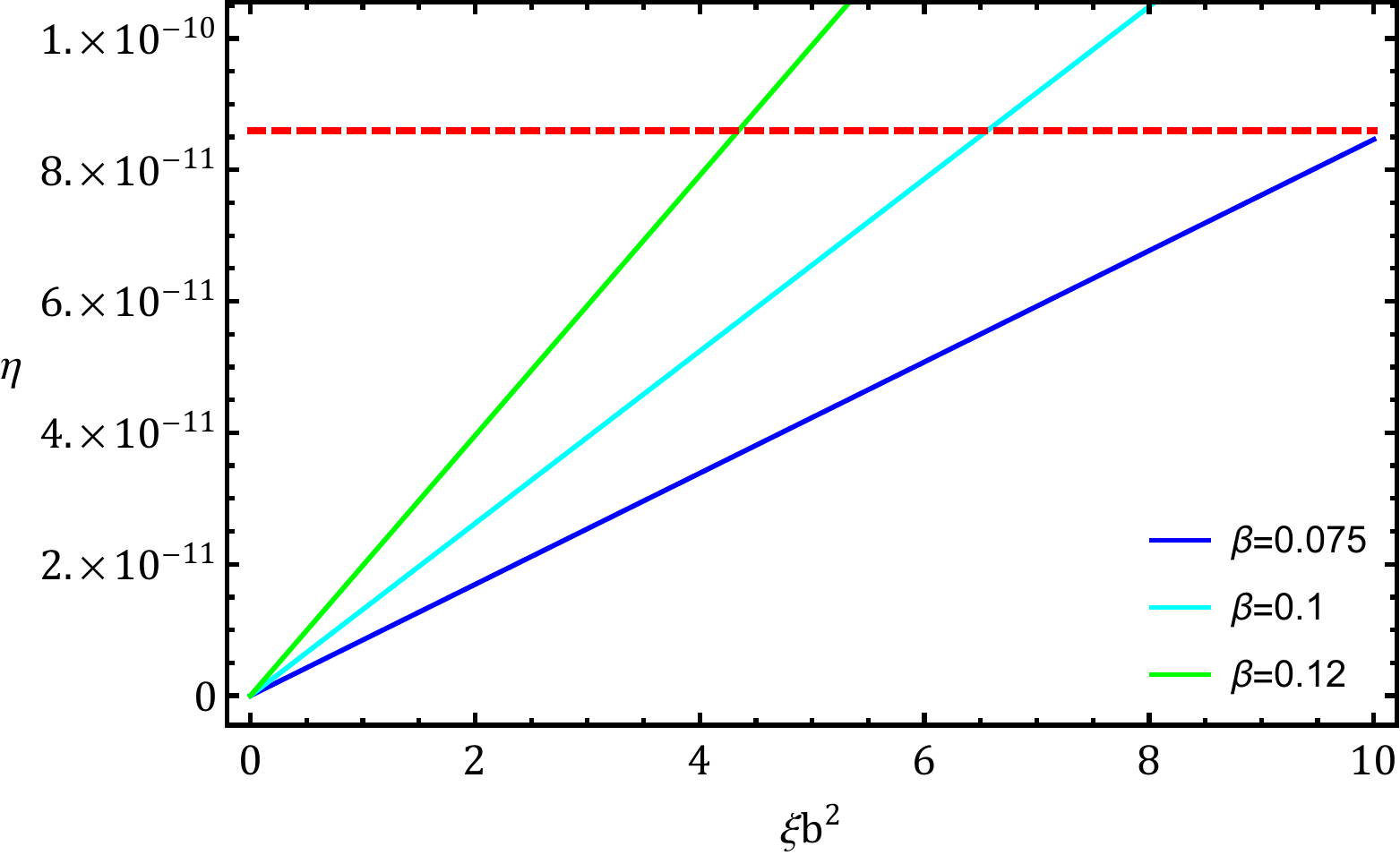}
		\end{tabular}
		\caption{Up row: Regions of existence in the $\beta-\xi b^2$ parameter space which satisfies the constraint (\ref{uper}). Bottom row: $\eta$ from  (\ref{basB}) in terms of $\xi b^2$ for optional values of $\beta$ which put in correspond allowed region. Here we set numerical values: ${\tilde M}=\bar{M}_P$= $M_*$, $T_D\sim10^{16}$GeV, $g_b=2$ and $g_*\sim107$.}
		\label{BA1}
	\end{figure*}
	
	\begin{figure*}[ht!]
		\begin{tabular}{c}
			\includegraphics[scale=0.45]{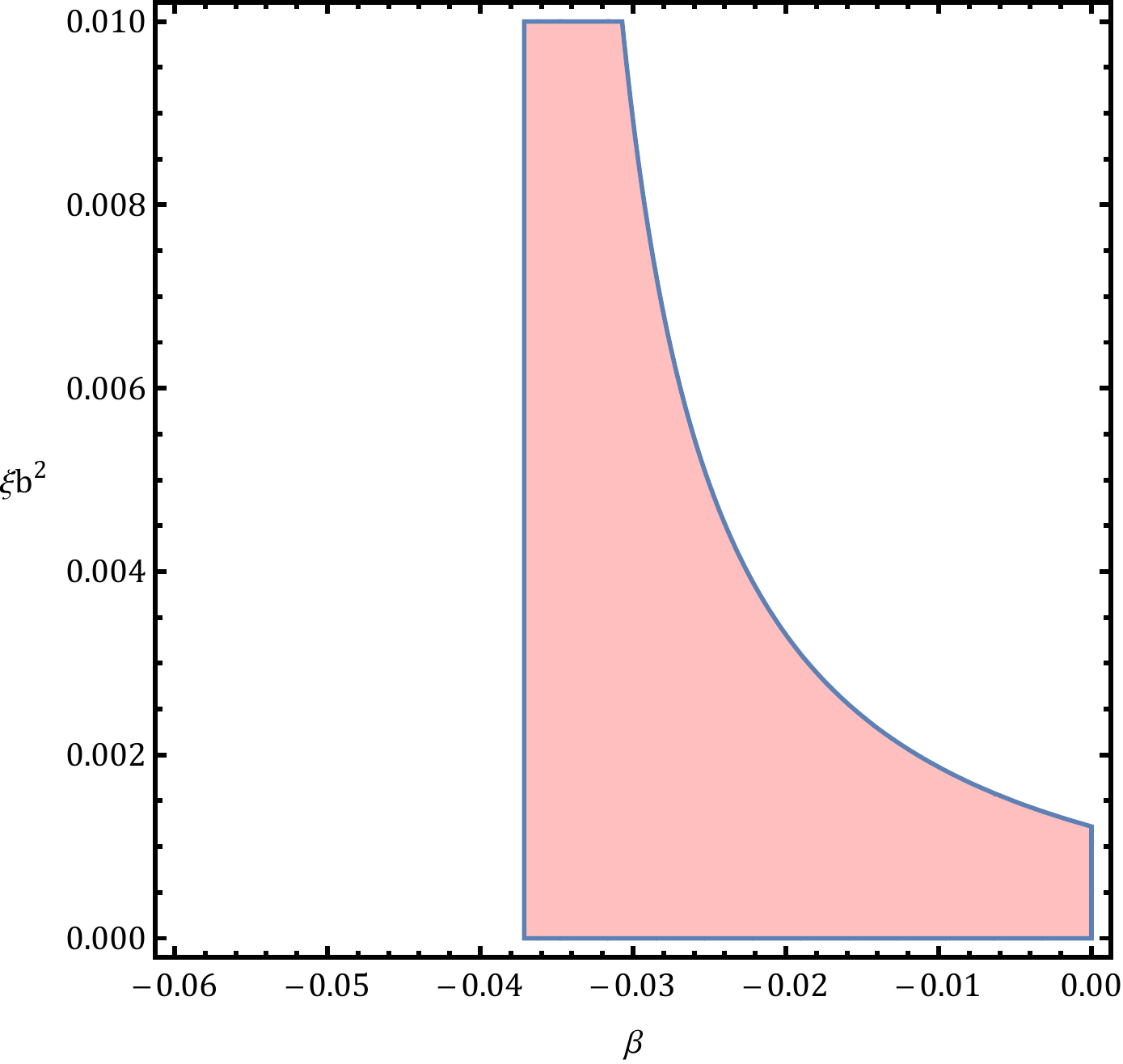}~~
			\includegraphics[scale=0.45]{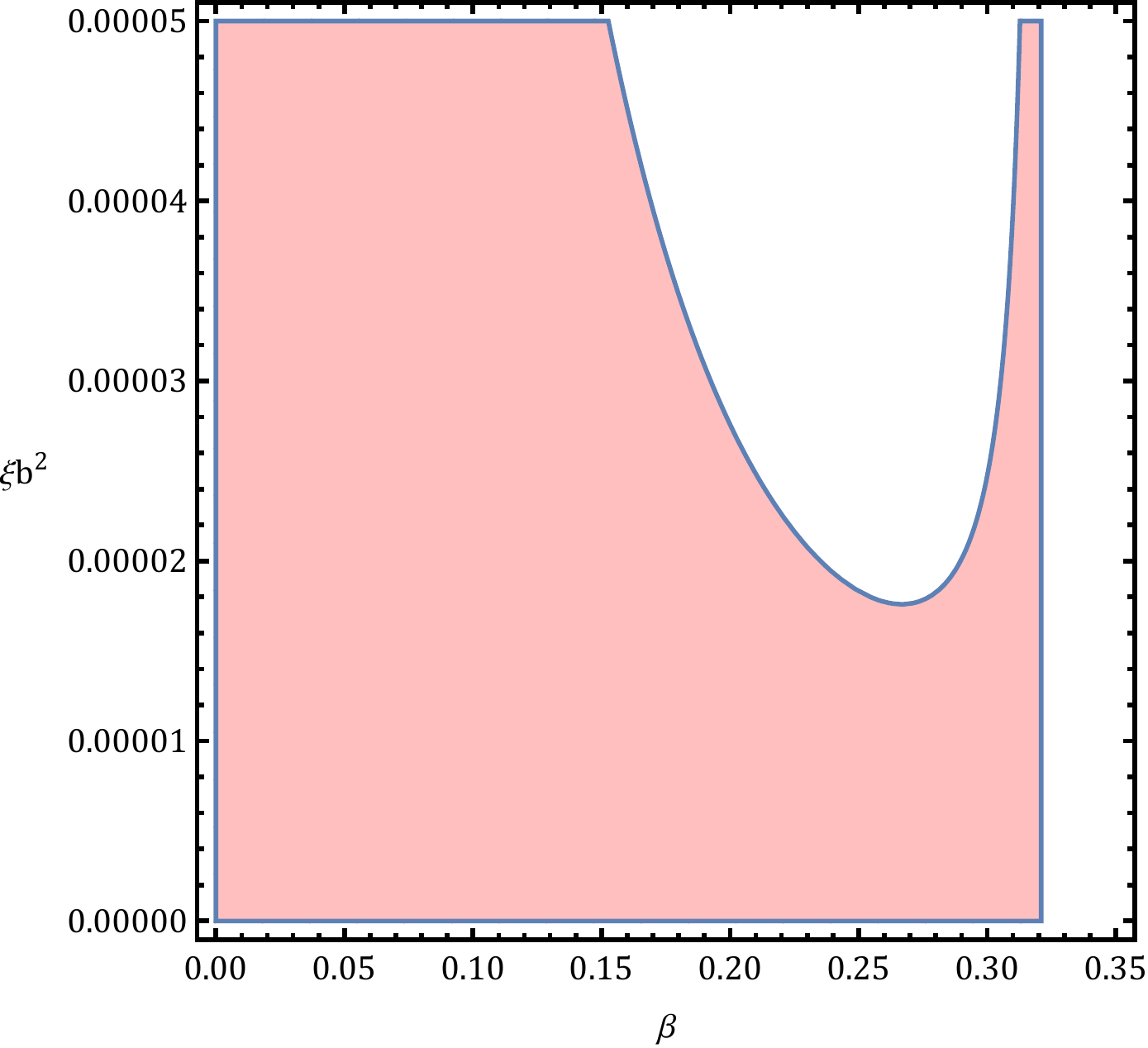}\\
			\includegraphics[scale=0.45]{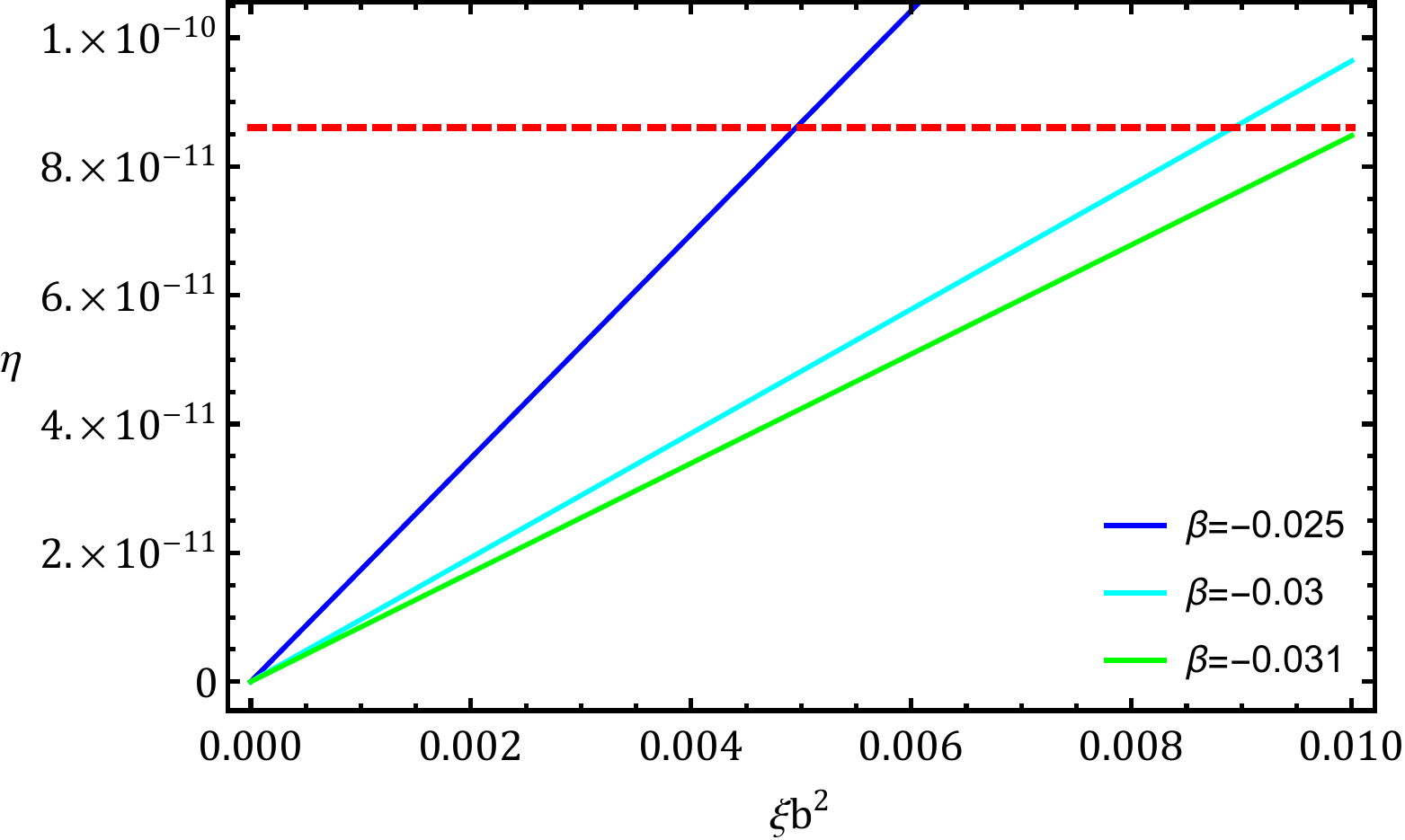}~~
			\includegraphics[scale=0.45]{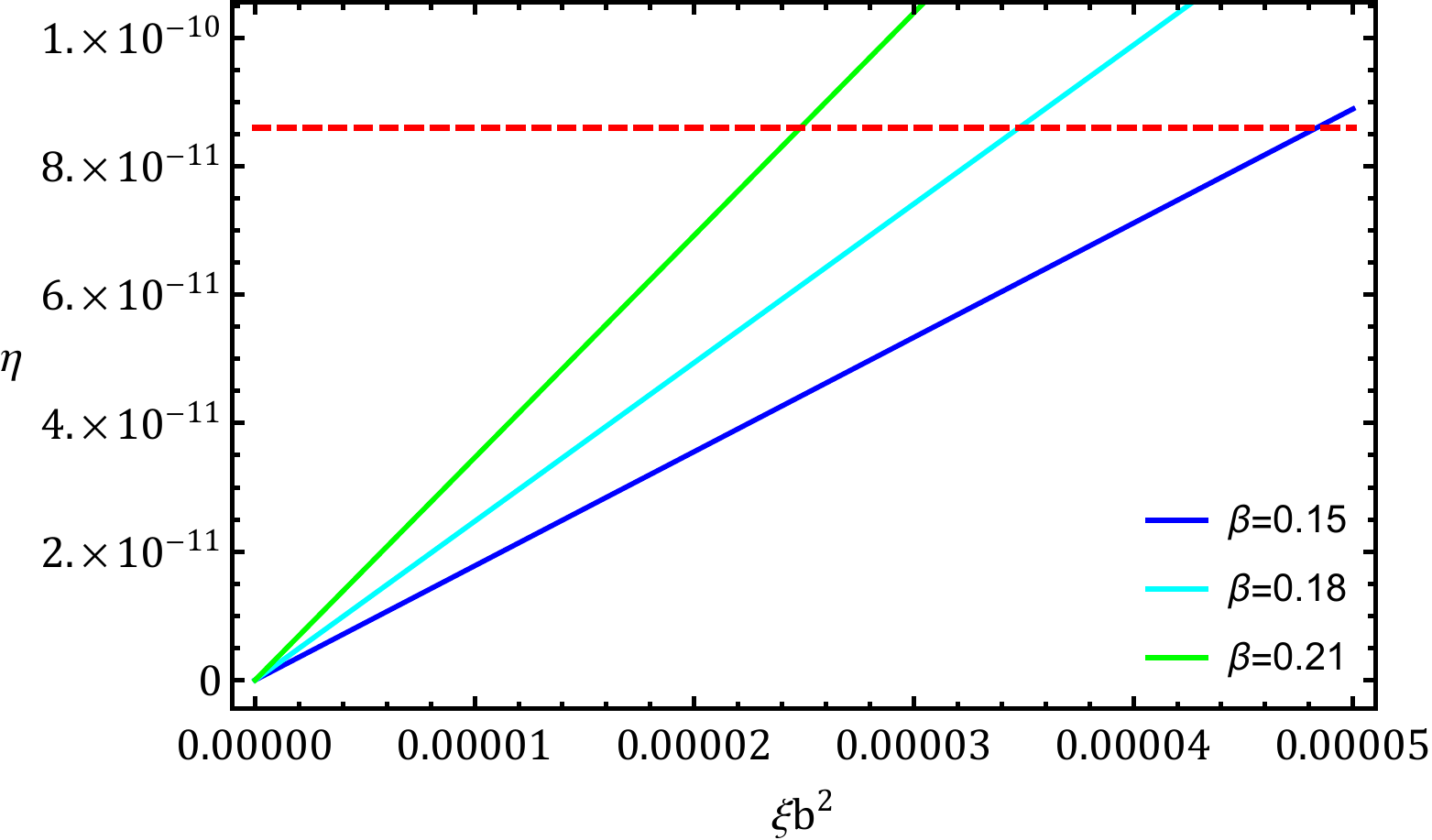}
		\end{tabular}
		\caption{Same as the Fig.~(\ref{BA1}) but for $M_*\sim10^{16}$GeV i.e., around GUT scale.}
		\label{BA2}
	\end{figure*}
	
	\begin{figure*}[ht!]
		\begin{tabular}{c}
			\includegraphics[scale=0.45]{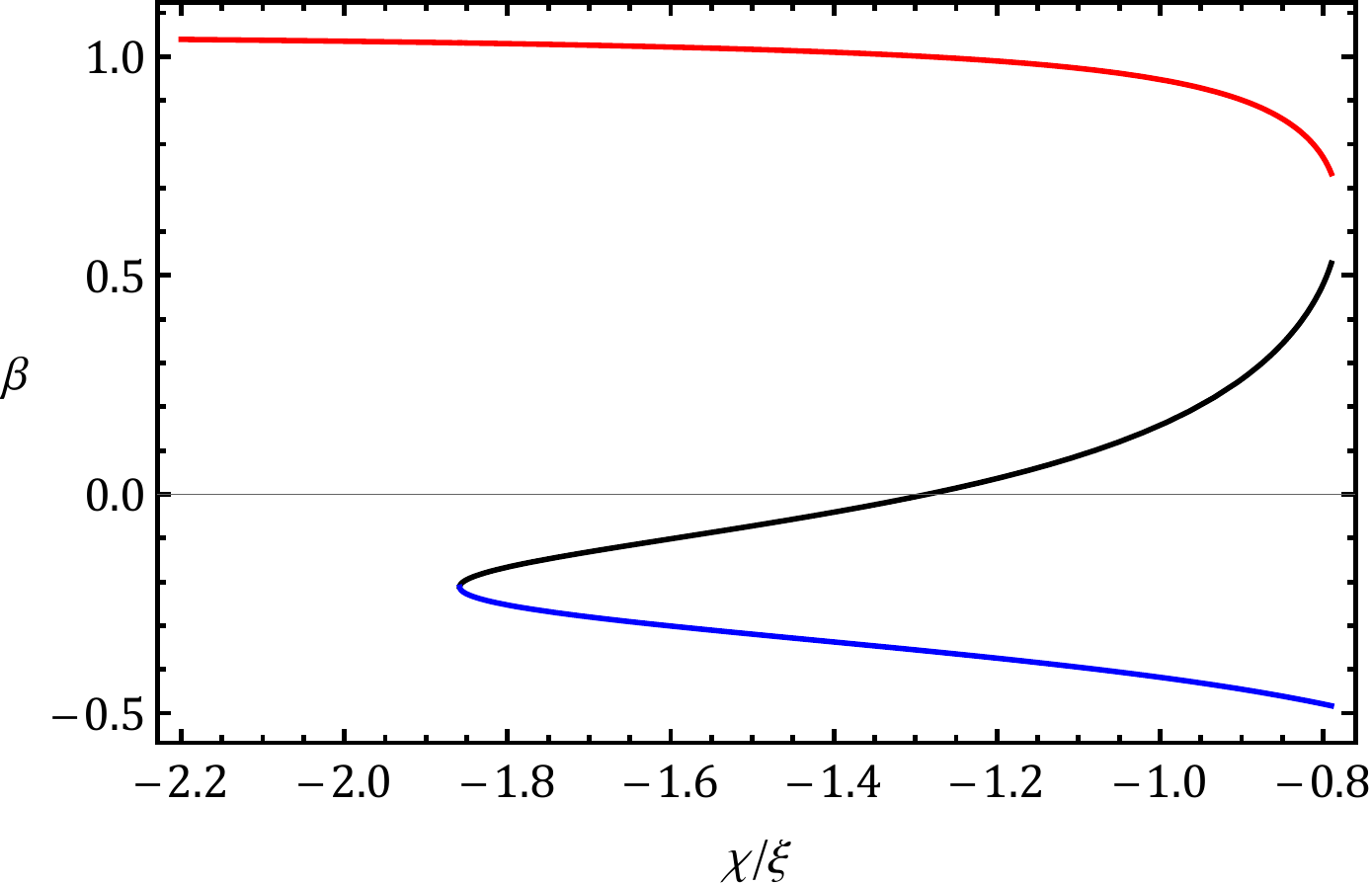}~~
			\includegraphics[scale=0.45]{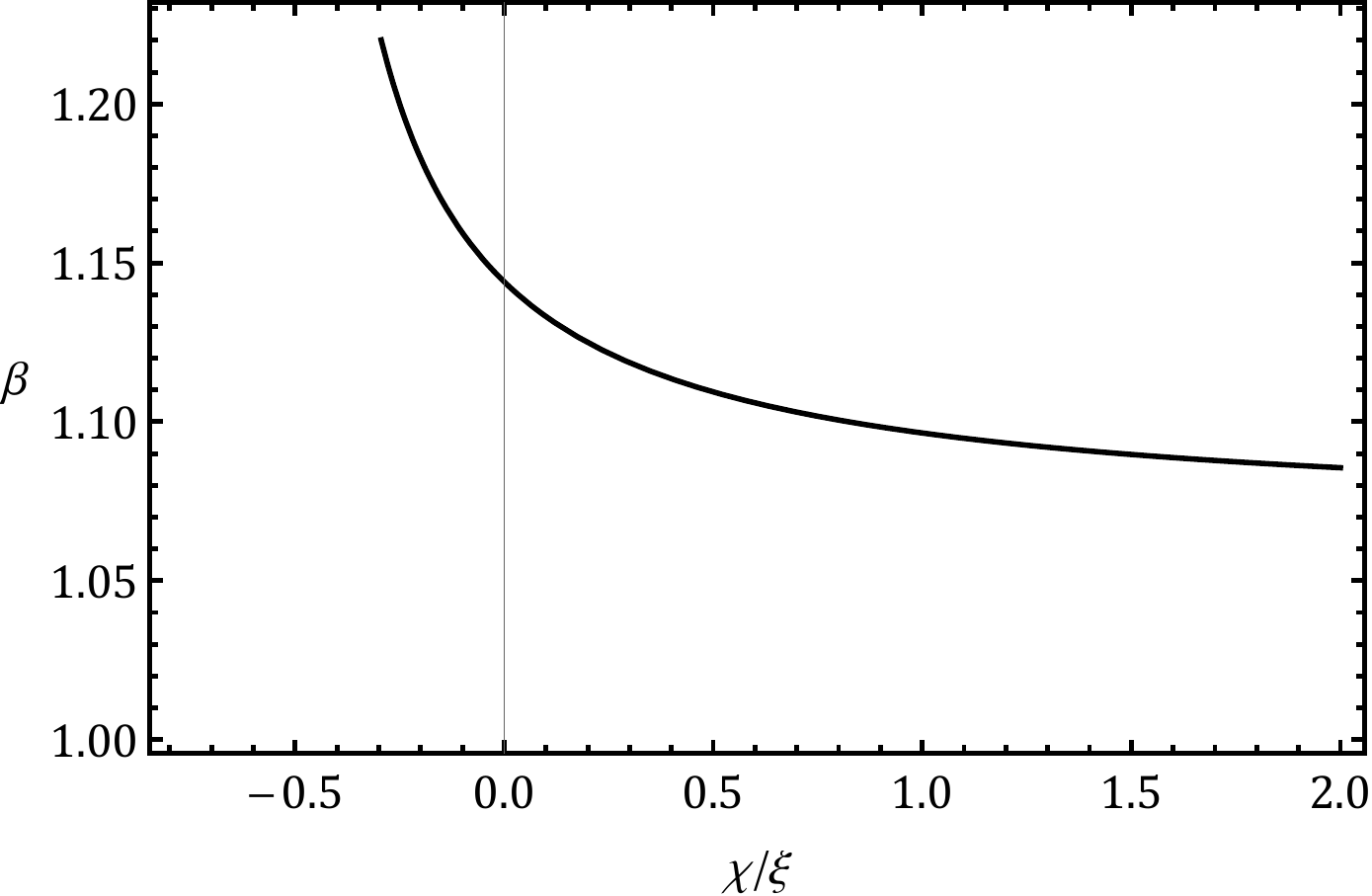}\\
		\end{tabular}
		\caption{The behavior of exponent $\beta$ in terms of dimensionless ratio $\frac{\chi}{\xi}$ for RD epoch. The left and right panels dedicate to cases $\frac{\chi}{\xi}\leq-0.8$ with three real solutions (black, blue, and red), and $\frac{\chi}{\xi}>-0.8$ with just one real solution, respectively.}
		\label{beta}
	\end{figure*}
	
	\begin{figure*}[ht!]
		\begin{tabular}{c}
			\includegraphics[scale=0.38]{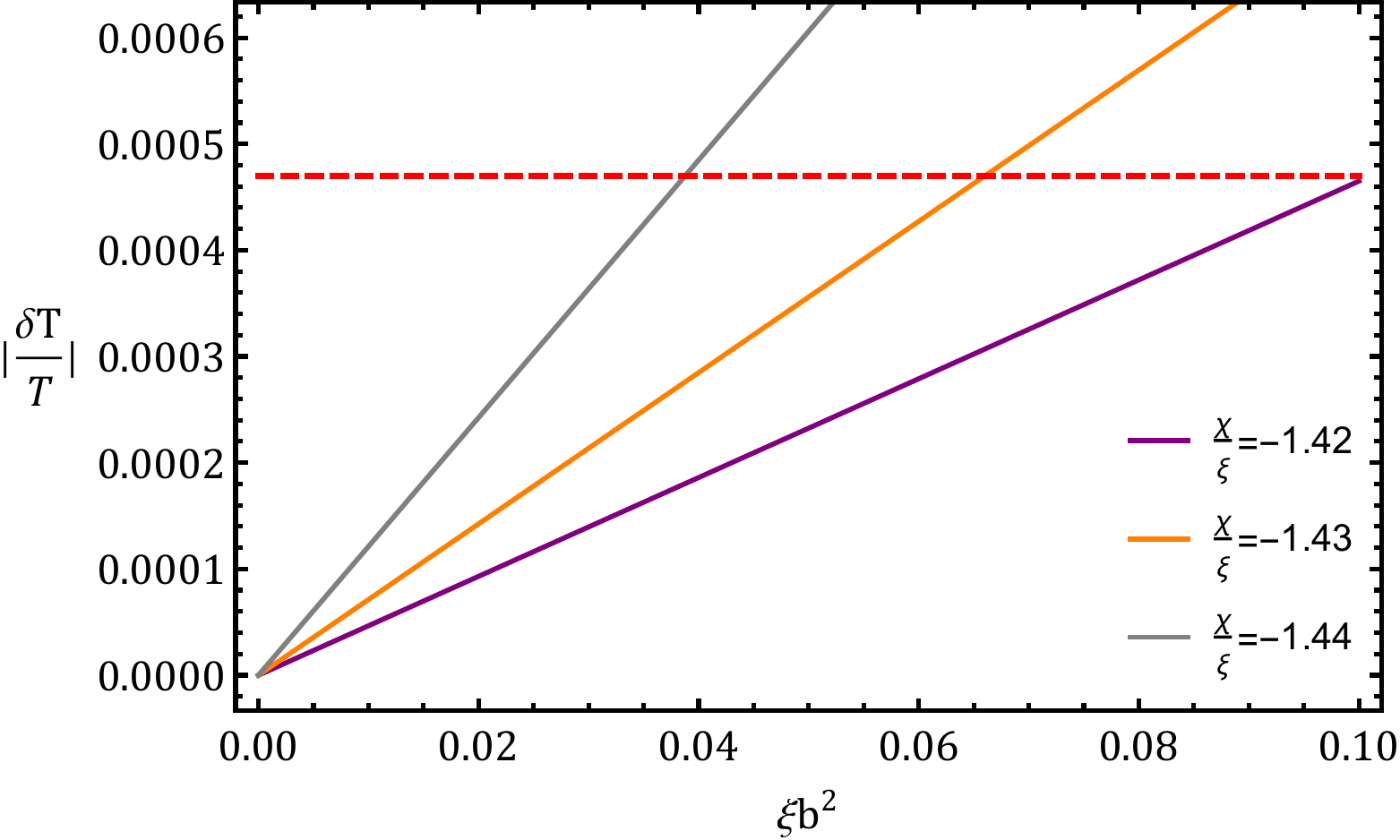}
			\includegraphics[scale=0.38]{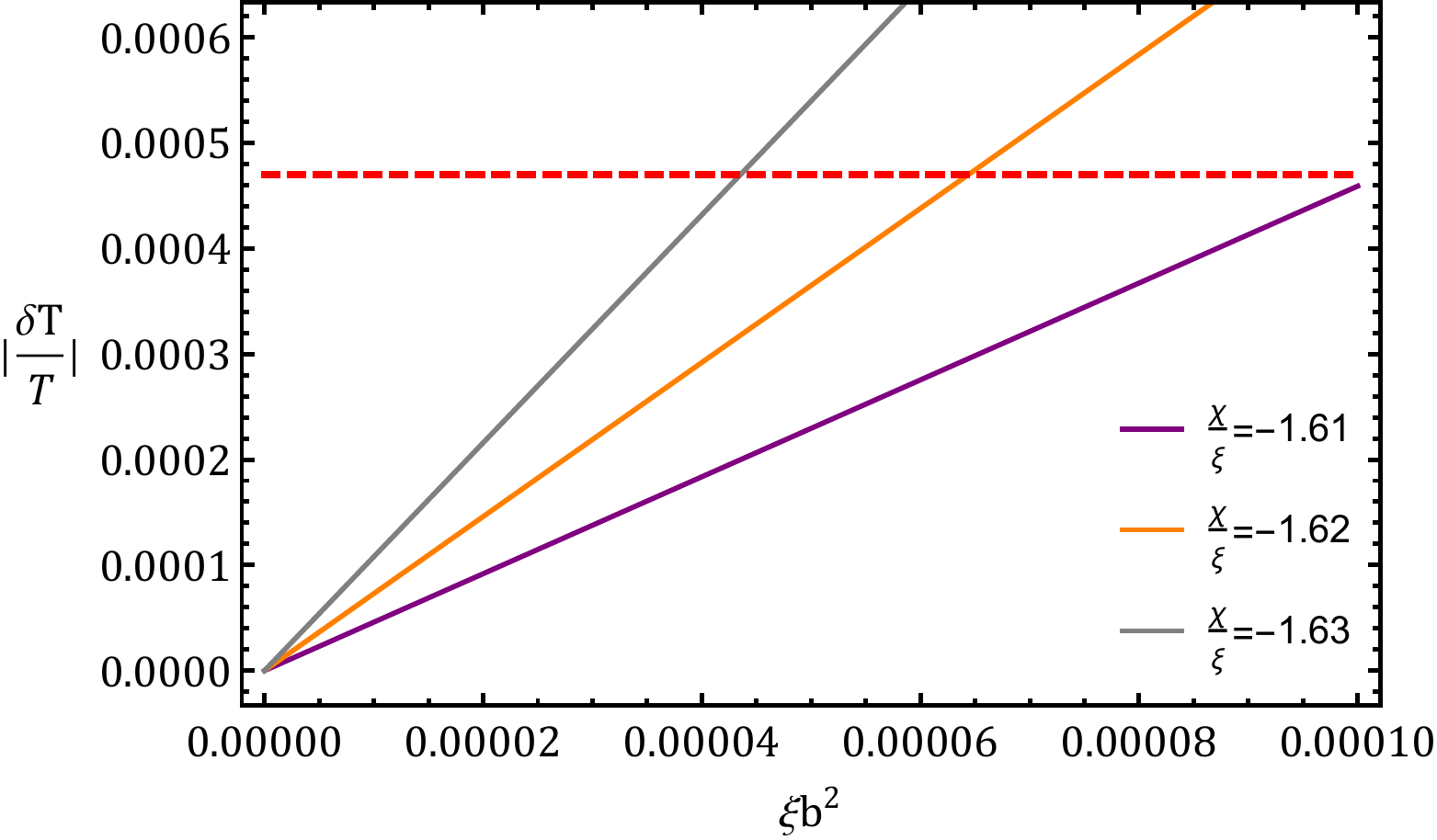}
			\includegraphics[scale=0.38]{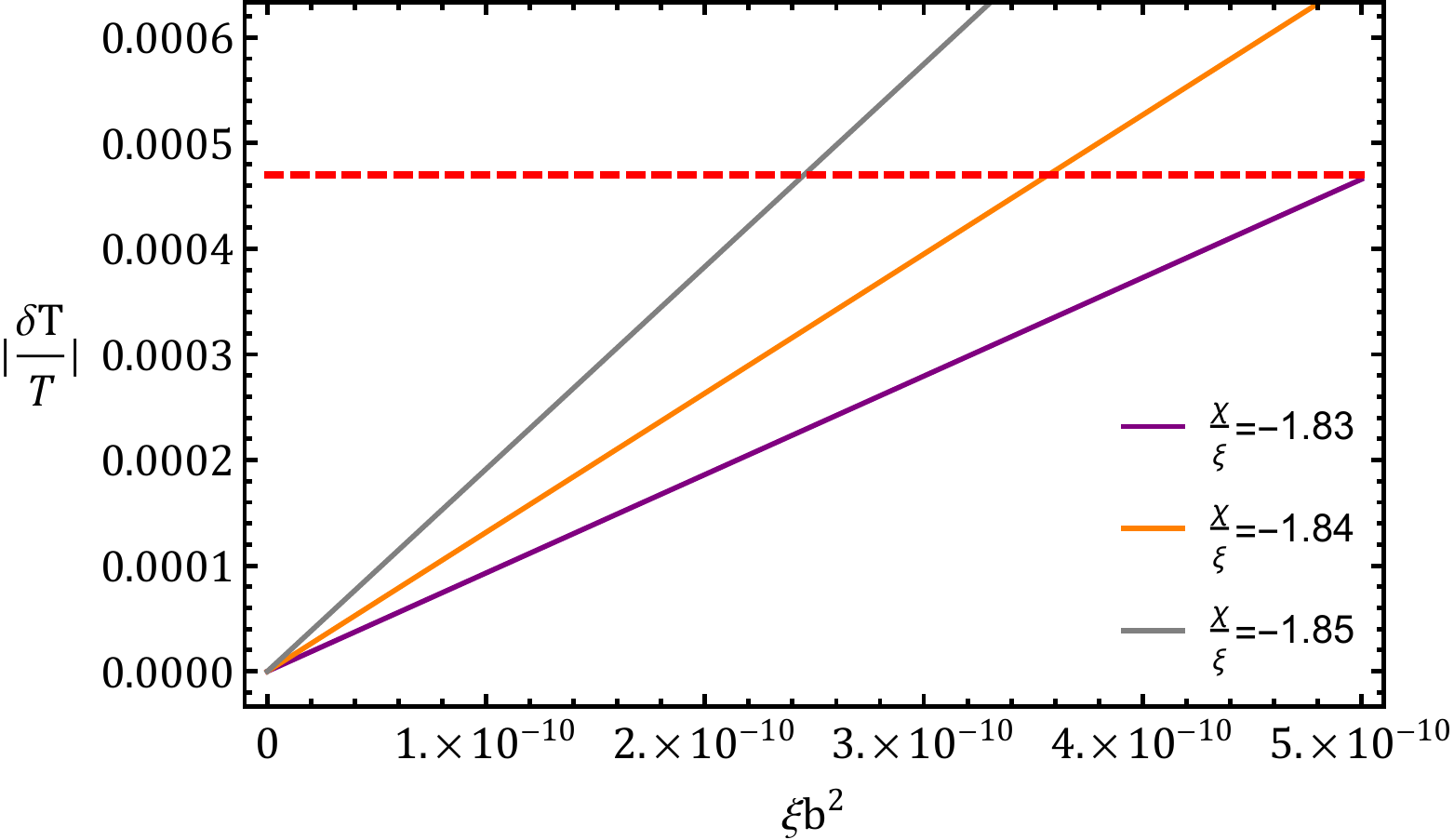}
		\end{tabular}
		\caption{$\left|\frac{\delta {T}_f}{{T}_f}\right|$ from  (\ref{deltaTTboundG}) in terms of $\xi b^2$ for some selected values of $\frac{\chi}{\xi}$ within the range $-1.85\leq\frac{\chi}{\xi}<-1.2$. 
			The rest of the numerical values are equal to the same values in Fig. \ref{NNB1}.}
		\label{RR}
	\end{figure*}
	\begin{figure*}[ht!]
		\begin{tabular}{c}
			\includegraphics[scale=0.45]{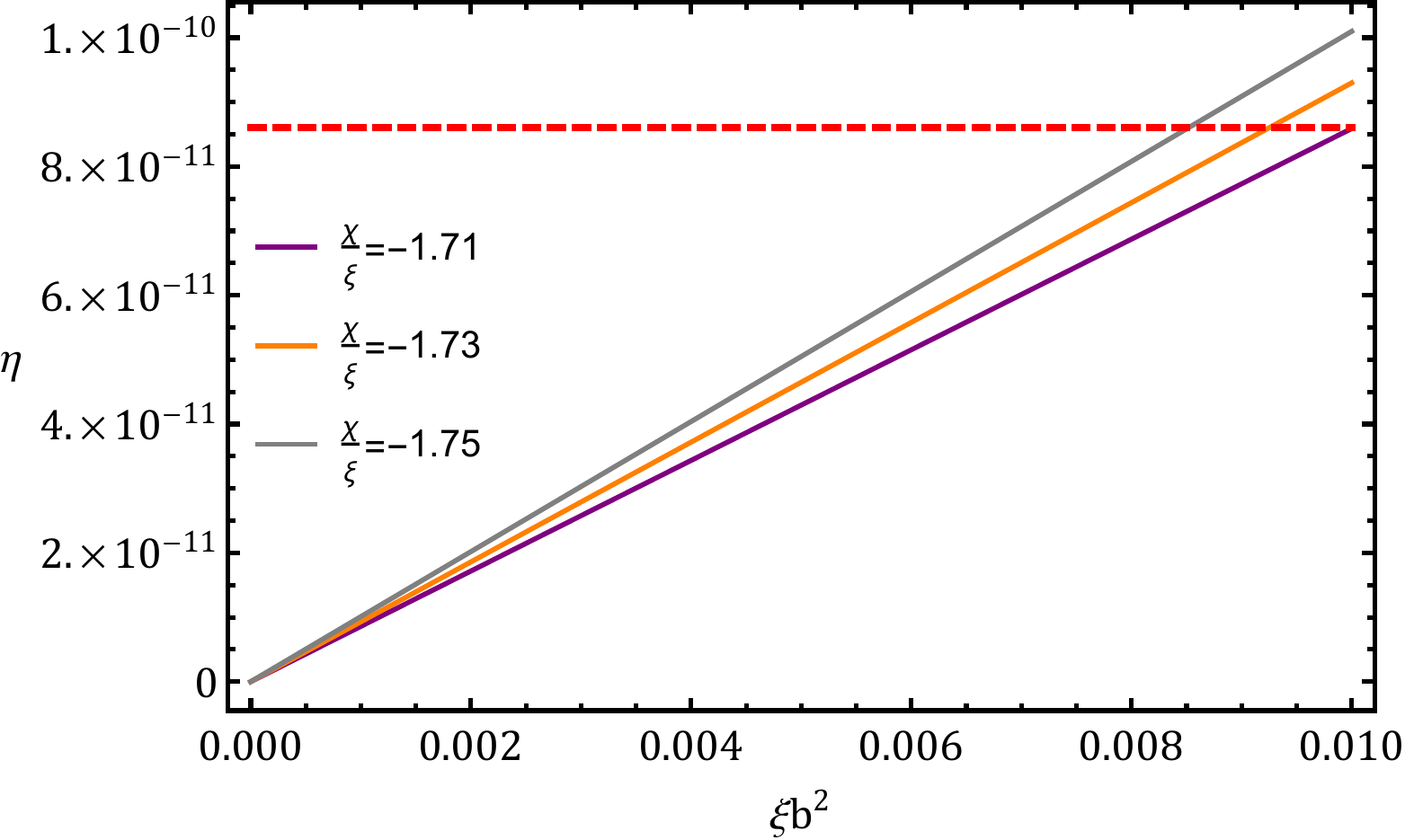}~~
			\includegraphics[scale=0.45]{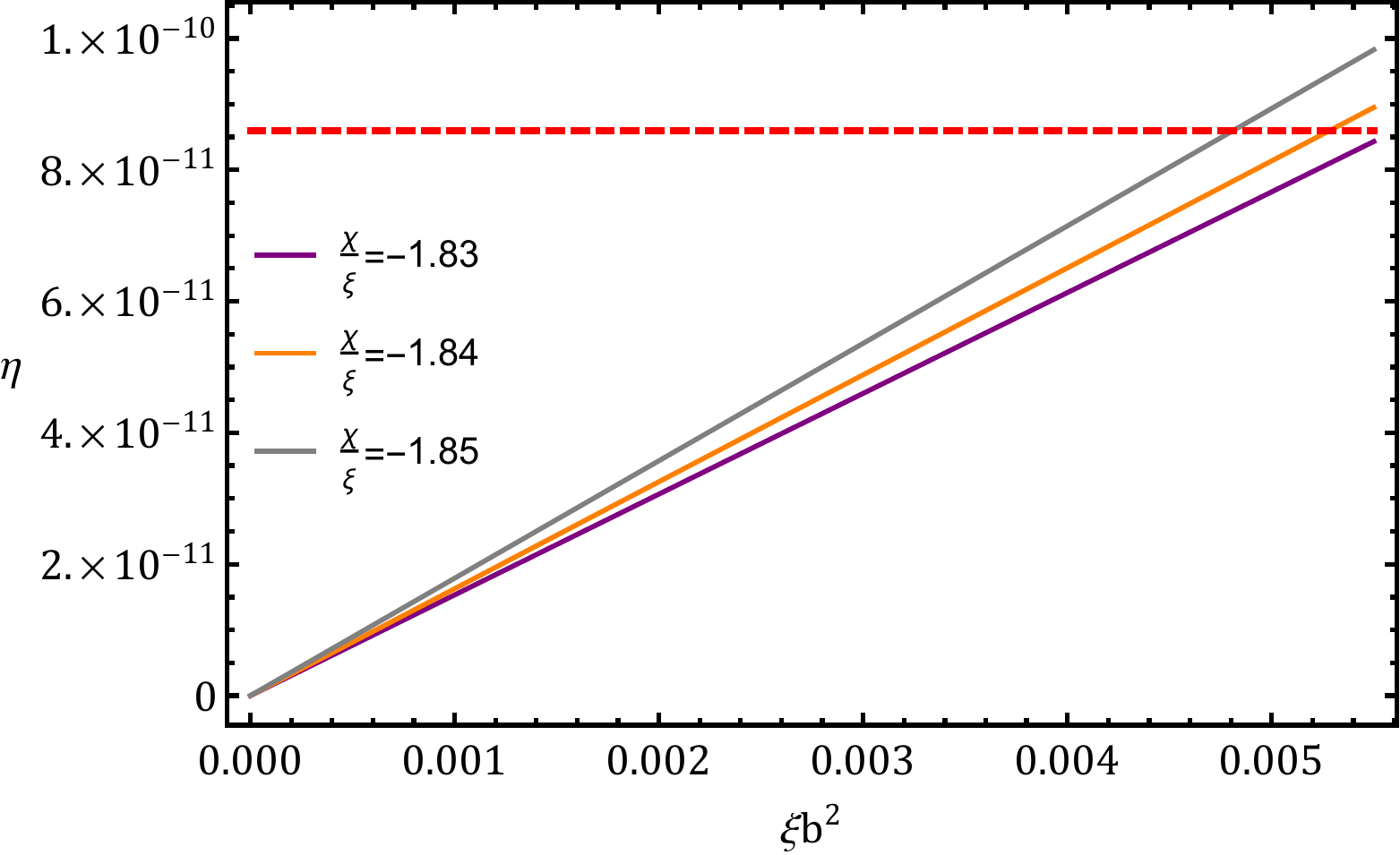}
		\end{tabular}
		\caption{$\eta$ from  (\ref{basB}) in terms of $\xi b^2$ for some selected values of $\frac{\chi}{\xi}$ within the range $-1.85\leq\frac{\chi}{\xi}<-1.2$. The rest of numerical values are equal to the same values in Fig. \ref{BA2}.}
		\label{RRR}
	\end{figure*}
	
	\section{Gravitational baryogenesis in Bumblebee cosmology}
	\label{Bar}
	In the light of supergravity theories there exist a mechanism for
	inducing baryon asymmetry during the evolution of the Universe,
	which has been proposed in \cite{KU1,KU2}. In this model, the
	thermal equilibrium is preserved, so that not all of Sakharov's
	conditions are fulfilled. The interaction responsible for the
	(dynamical) CPT violation is given by \cite{gravbar}
	\begin{eqnarray}\label{intterm}
		\frac{1}{M_*^{2}}\int\mathrm{d}^4x\sqrt{-g}\,J^\mu\partial_\mu R\,,
	\end{eqnarray}
	where $M_*$ is the cutoff scale characterizing the effective
	theory (typically it is of order reduced Planck mass
	$\bar{M}_P\sim2.4 \times 10^{18}$GeV),  and $J^\mu$ the baron
	current\footnote{Notice that $J^{\mu}$ can be any current leading
		to a net $B-L$ charge in equilibrium ($B, L$ are the baryon/lepton
		number) so that the asymmetry is not wiped out by the electroweak	anomaly \cite{krs1985}.} (see Refs.
	\cite{Lambiase:2006dq,Lambiase:2006ft,allGB1,allGB2,allGB3,allGB4,allGB5,allGB6,allGB7,allGB8,allGB9}
	for further applications). In the vacuum, the interaction
	(\ref{intterm}) violates $CP$, while $CPT$ is conserved. In an
	expanding universe, the interaction (\ref{intterm}) dynamically
	breaks $CPT$, generating an energy shift that is responsible for
	the asymmetry between particles and antiparticles. Moreover, the
	existence of interactions that violate baryon processes in thermal
	equilibrium is essential so that a net baryon asymmetry can be
	generated and gets frozen at the decoupling temperature $T_D$. It,
	in essence, is the temperature at which the baryon asymmetry
	generating interactions happen and due to the fact that the
	expansion rate of the Universe is larger than the interaction
	rate, it remains fixed since the interaction is less frequent.

	In an expanding universe, when the temperature drops below $T_D$,
	Eq. (\ref{intterm}) conducts us to the following relation
	\begin{eqnarray}\label{integrand}
		\frac{1}{M_*^2}J^\mu\partial_\mu R=\frac{1}{M_*^2}(n_B-n_{\Bar{B}})\dot{R}~,
	\end{eqnarray}
	where ${\dot R}$, $n_B$, and $n_{\Bar{B}}$ denote the time derivative of the Ricci scalar, baryon and anti-baryon number density, respectively.
	This relation allows defining the  {\it effective chemical potential} for baryons $\mu_B$, and for anti-baryons $\mu_{\Bar{B}}$, so that
	\begin{eqnarray}
		\label{chempot}
		\mu_B=-\mu_{\Bar{B}}=-\frac{\dot{R}}{M_*^2}~,
	\end{eqnarray}
	since (\ref{integrand}) corresponds to the energy density term for a grand canonical ensemble. For relativistic particles, the net  baryon number density reads \cite{kolb}
	\begin{eqnarray}
		n_B-n_{\Bar{B}}=\frac{g_b}{6}\mu_BT^2~,
	\end{eqnarray}
	where $g_b\sim\mathcal{O}(1)$ is the number of intrinsic degrees of freedom of baryons.
	The above relations allow writing the parameter $\eta$ characterizing the baryon asymmetry in the following form \cite{kolb}
	\begin{eqnarray}\label{base}
		\eta\equiv\frac{n_B-n_{\Bar{B}}}{s}\approx\frac{n_B}{s}\simeq-\frac{15\,g_b}{4\pi^2g_*}\frac{\dot{R}}{M_*^2T}\bigg|_{T_D}~,
		\label{asym1}
	\end{eqnarray}
	where $s=\frac{2\pi^2g_{*s}}{45}T^3$ is the entropy density (in
	the radiation-dominated era), and $g_{*s} \sim g_* \sim 107$ (here
	$g_{*s}$ is the number of degrees of freedom for particles which
	contribute to the entropy of the Universe, while $g_*$ the total
	number of degrees of freedom of relativistic particles
	\cite{kolb}.
	
	As it arises from (\ref{base}),  $\eta$ is different from zero if
	$\dot{R}\neq 0$. As we are going to discuss, the presence of the
	bumblebee vector field in the background break thermal equilibrium
	and modifies $\dot{R}$, making it non-vanishing so that $\eta \neq
	0$.
	
	By reminding of GR, the Ricci scalar $\dot{R}$ is computed by the trace of Einstein field equations so that one gets $R=-\kappa\,{\cal T}=-\kappa\,(\rho-3p)$ (see Eq. (\ref{TraceEq1})).
	In particular, during the radiation-dominated era, in which we are interested, the trace ${\cal T}$ vanishes (since the adiabatic index is $w=1/3$), meaning that $R=0$, and no net baryon asymmetry can be generated $\eta \sim {\dot R}=0$. This conclusion changes in the presence of a bumblebee background vector field. Actually, in such a case, the total energy-momentum is given by radiation and the bumblebee field $B$, so that the total trace does not vanish.
	As a results, Eq. (\ref{TraceEq}) reads off
	\begin{equation}\label{ricciexp}
		R = - T^{(B)}\,,   \end{equation}
	where $T^{(B)}$ is given in (\ref{TBFRW}),
	so that
	\begin{equation}\label{dtricci}
		\dot{R}= -6 N_{\xi, \chi} \xi b^2 {\tilde M}^3\left( {\tilde M} \, t\right)^{2\beta-3} \,,
	\end{equation}
	with
	\begin{eqnarray}	\label{Nxichi}
		N_{\xi, \chi} &=& \left(1+\frac{\chi}{\xi}\right)(\beta-1) (3\alpha^2+4\alpha\beta-\alpha) + \\
		&+&
		\left(1+2\frac{\chi}{\xi}\right)  
		\left(2\beta^2+\alpha \beta-\beta\right)+ \nonumber \\
		&+& 2\alpha\beta\left((2\alpha^2+2\alpha-2)\frac{\chi}{\xi}+\alpha-1\right)\,. \nonumber 
	\end{eqnarray}
	Note that the above equations are obtained by using Eq. (\ref{dotV}).
	We recall that  during the RD era the cosmic time $t$ and the temperature $T$ are related as
	\begin{equation}\label{tvsTGR}
		{\displaystyle \frac{1}{t}\simeq (\frac{4\pi^2 g_*}{90})^{1/2}\frac{{
					T}^2}{\bar{M}_{P}}}\,,
	\end{equation}
	or, equivalently ${T}(t)\simeq (t/\text{sec})^{-1/2}$MeV (notice
	that the entropy conservation $S\sim a^3 {T}^3=constant$ implies
	$T(t) a(t)=T_0 a_0$, where $T_0$ and $a_0$ are the temperature and
	scale factor of the Universe, $T_0\simeq 10^{-4}$eV, $a_0=1$,
	respectively). By substituting (\ref{dtricci}) in the baryon
	asymmetry formula (\ref{base}), one obtains
	\begin{equation}\label{basB}
		\eta = \varpi \, {N}_{\xi, \chi}  \xi b^2\, \left(\frac{\tilde M}{M_*}\right)^2 \left(\frac{T_D}{\bar{M}_P}\right)^{3-2\beta}\left(\frac{T_D}{{\tilde M}}\right)^{2-2\beta}\,,
	\end{equation}
	where the constant $\varpi$ is given by
	\[
	\varpi \equiv g_b\left(\frac{2\pi^2 g_*}{45}\right)^{(1-2\beta)/2} \,.
	\]
	Now, using the bound on the baryon asymmetry parameter
	$\eta$, that is \cite{ParticleDataGroup:2018ovx,AharonyShapira:2021ize}
	\begin{equation}\label{uper}
		0<\eta \lesssim \eta_{obs}\sim 8.6 \times 10^{-11}\,,
	\end{equation}
	we can extract some explicit constraints on $\xi b^2$ in interplay with negative and positive values of exponent parameter $\beta$.

	Now, by imposing the above-mentioned constraint on Eq. (\ref{basB}), in Fig. \ref{BA1} (up row) we illustrate the parameter space plots in terms of $\beta-\xi b^2$ which address the allowed regions in which the upper bound (\ref{uper}) is satisfied. Also, in the bottom row of this figure, we draw the plots $\eta-\xi b^2$ for some values of $\beta$ which are put in the corresponding allowed region. Concerning the case $\beta<0$ we find that independent of value of $\xi b^2$, for $\beta\leq-0.038$, the baryon asymmetry parameter becomes negative which is meaningless and not acceptable. So, by adopting range $-0.038<\beta<0$, one can extract some upper bounds around $10^{2-3}$ for $\xi b^2$. As one can see, by going to the case $\beta>0$ this upper bound will improve a few orders of magnitude.
	
	An interesting result we found here is that the cutoff scale $M_*$
	plays an inevitable role in falling the upper bounds on $\xi b^2$.
	As we can see from Fig. \ref{BA1} there we have fixed $M_*=\bar{M}_P$,
	corresponding to the Planck scales at which the interaction
	(\ref{base}) is effective. In essence, we deal with an effective
	theory, and fix $M_*$ a few orders of magnitude
	lower e.g., around the GUT scale ($M_*\sim10^{16}$GeV). In this
	case, the upper bounds released for $\xi b^2$ in Fig. \ref{BA1} improve a
	few orders of magnitude, see Fig. \ref{BA2}. Despite
	these improvements, in comparison with upper bounds extracted from
	BBN in the previous section, we still do not deal with stringent
	constraints on $\xi b^2$. Indeed, the achievement worth of noting
	here is not related to the upper bound derived for $\xi b^2$, but
	is for restricting the evolution rate of the bumblebee vector
	field i.e., $-0.038<\beta<0$. The worth of this constraint is to
	consider it complementary to BBN, in the sense that other values
	belonging to the $\beta\leq-0.038$ range in BBN analysis are ruled
	out. In this way, the most conservative constraint extracted
	within the range  $-0.038<\beta<0$ for the VEV of the bumblebee
	timelike vector field i.e., $\xi b^2$ is $\lesssim 10^{-24}$. We
	say the most conservative since for all values except for
	$\beta=-0.038$ within the allowed range of $\beta$, the
	above-mentioned upper bound gets tighter, as one can see of Fig.
	\ref{NNB1} (the right panel in the bottom row).
	
	\section{New strategy: Constraints on $\xi b^2$ in interplay with $\frac{\chi}{\xi}$}\label{New}
	
	So far, all constraints derived for $\xi b^2$ from BBN and Baryogenesis come, in essence,  from the interplay with exponent $\beta$. More exactly, by solving Eq. (\ref{CB=0}) in terms of $\frac{\chi}{\xi}$ for RD era, we indeed treated $\beta$ as a free parameter. The benefit of this approach is that it lets us probe $\xi b^2$ in explicit interplay with the free parameter $\beta$ related to the evolution of the bumblebee field in the RD era.  
		Alternatively, there is another possibility in which Eq. (\ref{CB=0}) is solved in terms of $\beta$ and, subsequently, the dimensionless ratio $\frac{\chi}{\xi}$ this time is treated as the involved parameter. This gives us the possibility of probing $\xi b^2$ in explicit interplay with $\frac{\chi}{\xi}$ as coupling constants in the action (\ref {eqAction}).
	
	By solving Eq. (\ref{CB=0}) for RD era in terms of $\beta$, we have a cubic equation such as
		\begin{equation}\label{Cu}
			\beta^3-\bigg(\frac{7+18\frac{\chi}{\xi}}{16+32\frac{\chi}{\xi}}\bigg)\beta^2-\bigg(
			\frac{9+14\frac{\chi}{\xi}}{16+32\frac{\chi}{\xi}}
			\bigg)\beta-\bigg(
			\frac{9+7\frac{\chi}{\xi}}{32+64\frac{\chi}{\xi}}\bigg)=0.
		\end{equation}
		It is not difficult to show that the cubic equation above for $\frac{\chi}{\xi}\leq-0.8$ has three real solutions, while it has just one real solution for $\frac{\chi}{\xi}>-0.8$. We display both cases in Fig. \ref{beta}. It is observed that just two solutions marked with black and blue curves in the left panel, address $\beta<0$ (as the desirable case of cosmology, as already stated). The range $-1.85\leq\frac{\chi}{\xi}<-1.2$ is the common region of $\frac{\chi}{\xi}$ for these two solutions. 
		Now, by taking into account the solution marked with the blue curve in Fig. \ref{beta}  and using the BBN constraint  (\ref{deltaTTboundG}), we can plot $\left|\frac{\delta {T}_f}{{T}_f}\right|$ in terms of $\xi b^2$ for values of $\frac{\chi}{\xi}$ within the aforementioned range, see Fig \ref{RR}. It can be seen that the upper bound of $\xi b^2$ moves from $10^{-2}$ to $10^{-10}$, as the value of the dimensionless ratio $\frac{\chi}{\xi }$ approaches its extreme one i.e., $-1.85$. So, it is easy to recognize that for values $-1.2\leq\frac{\chi}{\xi}<-0.8$, we will deal with the resulting very weak upper bounds for $\xi b^2$.
	
	In this regard, by putting the favored solution of $\beta$ (corresponding to the blue curve in Fig. \ref{beta}) into the baryon asymmetry parameter $\eta$ in Eq. (\ref{basB}), we display in Fig. \ref{RRR} the plot of $\eta-\xi b^2$ for some selecting values of $\frac{\chi}{\xi}$. Here, the best upper bound for $\xi b^2$, which is not better than the order of magnitude $10^{-3}$, extracts by setting $\frac{\chi}{\xi}$ around the extreme value.  
	
Now, one can compare quantitatively the upper bounds obtained here for $\xi b^2$ and those were derived in the two previous Sections. One can infer that despite the constraints obtained from Baryogenesis in both approaches having almost the same order of magnitude, for BBN the former approach is more efficient since results in deriving tighter constraints on $\xi b^2$.
	
	\section{Conclusions}\label{Con}
	In this paper, we have considered a vector extension of the standard
	cosmology known as the bumblebee model in which by keeping
	isotropy and homogeneity of the Universe, the Lorentz symmetry
	spontaneously breaks by coupling a background time-like bumblebee
	vector field to Ricci tensor and scalar. We have used the
	implication of this cosmology model at hand for the formation of
	light elements and baryon asymmetry in the early universe, namely
	on the Big Bang Nucleosynthesis (BBN) and Baryogenesis
	respectively. By taking into account of a time-depending ansatz
	$\sim t^{\beta}$ for the evolution of the bumblebee field $B(t)$
	with cosmic time, we in Sections \ref{BBN} and \ref{Bar} have extracted some upper bounds on the
	vacuum expectation value (VEV) of the bumblebee timelike vector
	field i.e. $\xi b^2$. By solving Eq. (\ref{CB=0}) in terms of the dimensionless ratio $\frac{\chi}{\xi}$, we have analyzed both possible
		negative and positive ranges of exponent parameter $\beta$, with
		particular attention to the former, since it addresses the
		diluting of the bumblebee field as the Universe evolves, which is
		favored from the view of cosmology. From the combination of BBN
	and Baryogenesis, we find that, for the favourite scenario of the time-depending
	bumblebee vector field with a negative exponent parameter, 
	the constraints are: $-0.038<\beta<0$, and $\xi b^2\lesssim10^{-24}$. It
	is important to note that the above upper bound on $\xi b^2$ is
	derived in the case of setting $\beta\approx-0.038$, so that by
	going to within the allowed range of $\beta$, the upper bound
	gets a few orders of magnitude tighter. 
	
At the end of our analysis (Section \ref{New}), we pursued the strategy of solving Eq. (\ref{CB=0}) in terms of $\beta$. It lets us probe $\xi b^2$ this time in explicit interplay with the ratio $\frac{\chi}{\xi}$ made by two coupling constants embedded in the action (\ref {eqAction}). We have repeated the same analysis done in Sections related to BBN, and Baryogenesis, and derived some upper bounds for  $\xi b^2$. The comparison of upper bounds in Section \ref{New} with previous counterparts openly shows that the most stringent constraints for $\xi b^2$ come from the primary strategy in Sections \ref{BBN} and \ref{Bar}.
	
Referring to \ref{BM}, in particular to the connection between the exponent parameter $\beta$ and the general power-law form of the bumblebee potential $\left( B^\mu B_\mu \pm b^2 \right)^n$, there is a relation given by $\beta=\frac{-1}{n-1}$. As a consequence, the tight constraint $-0.038<\beta<0$ implies that the power-law bumblebee potential of the form $\left( B^\mu B_\mu \pm b^2 \right)^{n\leq 27}$ is ruled out. Concerning the new strategy, we saw that the favorite solution of Eq. (\ref{CB=0}) i.e., the blue curve in the left panel of Fig. (\ref{beta}), restricts the exponent parameter within the range $-0.47<\beta\leq-0.22$, corresponding to $3.12<n\leq5.5$. Overall, in light of both approaches, one should no longer worry about the instability issue raised in \cite{Bluhm:2008yt} for the existing cosmological model. 
	
	Finally, it is worth mentioning the significance of the
	results. First of all, very stringent constraints derived for $\xi
	b^2$ from BBN indicate the size of Lorentz violation for the early
	Universe with the same course of evolution expected from standard
	cosmology. In other words, these constraints have been obtained
	provided that the BBN predictions are preserved. Second, unlike
	the standard cosmological model, by taking the BG model into
	account, the gravitational baryogenesis mechanism allows for explaining
	the matter-antimatter asymmetry in the Universe induced by the bumblebee field.

	\acknowledgments
	
	M.Kh and A.Sh, thank Shiraz University Research Council. GL thanks INFN for support. We would like to appreciate the anonymous referee for insightful comments that helped us improve the paper.
	
	\appendix
	
	\section{Useful formulas}
	
	In this Appendix, we report some useful formulas.
	For the power law dependence of the scale factor, $a(t)=a_0 t^\alpha$, one has
	
	\begin{equation}
		\frac{\dot a}{a}=\frac{\alpha}{t}\,, \quad \frac{\ddot a}{a}=\frac{\alpha(\alpha-1)}{t^2}\,, \quad
		\frac{\dddot a}{a}=\frac{\alpha(\alpha-1)(\alpha-2)}{t^3}\,.
	\end{equation}
	The components of the energy-momentum tensor of the bumblebee field are given by
	\begin{eqnarray}
		T^{(B)0}_{\quad\quad 0} &=&\kappa V+3 B^2 H\left( (\xi + \chi)H + (\xi+2\chi)\frac{\dot B}{B}\right) \nonumber \\
		&\equiv & \rho_B\,, \label{TB00} \\
		& & \nonumber \\
		T^{(B)i}_{\quad \quad j} &=& \Big\{\kappa V + B^2 \Big[(\xi+\chi)\left(H^2+\frac{\ddot a}{a}+4H \frac{\dot B}{B}\right)+ \nonumber \\
		& & + (\xi+2\chi)\left(\frac{{\dot B}^2}{B^2}+\frac{\ddot B}{B}\right)\Big]\Big\}\delta^i_j \nonumber\\
		& \equiv & - p_B \, \delta^i_j \,. \label{TBij}
	\end{eqnarray}
	
	Using (\ref{ansatz}) one gets that in a FRW universe $\rho_B$ and
	$p_B$, Eqs. (\ref{TB00}) and (\ref{TBij}), are given by
	\begin{eqnarray}
		\rho_B &=& \kappa V+3b_0^2 \alpha [(\xi+\chi)\alpha + (\xi+2\chi)\beta]\frac{{\tilde M}^{2\beta}}{t^{2-2\beta}}\,, \label{rhoBFRW} \\
		& & \nonumber \\
		p_B &=& -\kappa V -b_0^2 \Big[ (\xi+\chi)\alpha(2\alpha-1+4\beta)+ \label{pBFRW} \\
		& & + (\xi+2\chi)\beta(2\beta-1)\Big]\frac{{\tilde M}^{2\beta}}{t^{2-2\beta}}\,. \nonumber
	\end{eqnarray}
	Moreover,
	\[
	{\dot \rho}_B = b_0^2 3\alpha (B_B- A_B) \, \frac{{\tilde M}^{2\beta}}{t^{3-2\beta}}\,,
	\]
	where $A_B$ and $B_B$ are defined in (\ref{ABdef}) and (\ref{BBdef}).
	
	\section{Big Bang Nucleosynthesis}
	
	We shortly review the main features of BBN
	\cite{kolb,bernstein}.
	In the early universe, the primordial ${}^4He$ was formed at
	temperature ${T}\sim {\cal O}(1)$ MeV. The (relativistic) electron, positron,
	neutrinos and photons are in thermal equilibrium owing to the rapid collision. The interactions
	involved are $\nu_e+n \leftrightarrow  p+e^-$, $e^++n \leftrightarrow p + {\bar \nu}_e$ and $n \leftrightarrow p+e^- + {\bar \nu}_e$.
	The neutron abundance is computed via the conversion rate of protons into
	neutrons ($\lambda_{pn}$) and its inverse ($\lambda_{np}$)
	\begin{equation}\label{LambdaA}
		\Lambda({T})=\lambda_{np}({ T})+\lambda_{pn}({T})\,,
	\end{equation}
	where
	\begin{equation}\label{sumprocess}
		\lambda_{np}=\lambda_{n+\nu_e\to p+e^-}+\lambda_{n+e^+\to p+{\bar
				\nu}_e}+\lambda_{n\to p+e^- +
			{\bar \nu}_e}\,.
	\end{equation}
	The rates $\lambda_{np}$ and $\lambda_{pn}$ are related as
	$\lambda_{np}({T})=e^{-{\cal Q}/{T}}\lambda_{pn}({T})$, with ${\cal
		Q}=m_n-m_p$ the mass difference of neutron and proton.
	The interaction rate
	for the process $n+\nu_e\to p+e^-$ is
	\begin{equation}\label{rateproc1}
		d\lambda_{n+\nu_e\to p+e^-}= d\mu \, |\langle{\cal M}|^2\rangle W \,,
	\end{equation}
	where the various terms are defined as
	\begin{eqnarray}
		d\mu & \equiv &  \frac{d^3p_e}{(2\pi)^3 2E_e} \frac{d^3p_{\nu_e}}{(2\pi)^3
			2E_{\nu_e}}\frac{d^3p_
			p}{(2\pi)^3 2E_p}\,, \label{dmu} \\
		W &\equiv &(2\pi)^4 \delta^{(4)}({\cal P})n(E_{\nu_e})[1-n(E_e)]\,, \label{WA}\\
		{\cal P}  &   \equiv &  p_n+p_{\nu_e}-p_p-p_e\,,  \\
		{\cal M} &= &\left(\frac{g_w}{8M_W}\right)^2 [{\bar u}_p\Omega^\mu u_n][{\bar
			u}_e\Sigma_\mu v_{\nu_e}]\,, \label{M} \\
		\Omega^\mu &\equiv & \gamma^\mu(c_V-c_A \gamma^5)\,,
		\\
		\Sigma^\mu&
		\equiv&
		\gamma^\mu(1-\gamma^5)
		\,.
	\end{eqnarray}
	From Eq. (\ref{rateproc1}) one gets
	\begin{equation}\label{rateproc1fin}
		\lambda_{n+\nu_e\to p+e^-}=A \, {\cal T}^5 I_y\,,
	\end{equation}
	where $A\equiv \frac{g_V+3g_A}{2\pi^3}$
	and
	\begin{equation}
		I_y=\int_y^\infty \epsilon(\epsilon-{\cal Q}')^2\sqrt{\epsilon^2-y^2}\, n(\epsilon-{\cal
			Q})[1-n(\epsilon)]d\epsilon,
	\end{equation}
	with $y\equiv \frac{m_e}{{T}}$ and ${\cal Q}'=\frac{{\cal Q}}{{T}}$.
	In a similar way, for the process $e^+ + n \to p+ {\bar \nu}_e$, one gets
	%
	%
	\begin{equation}
		\label{ne-pnu-fin}
		\lambda_{e^+ + n\to p+{\bar \nu}_e}=A\, {T}^3(4! {T}^2+2\times 3! {\cal Q}{T}+2! {\cal
			Q}^2)\,.
	\end{equation}
	Finally, the neutron decay follows from  $n\to p+e^- +{\bar \nu}_e$, giving
	\begin{equation}\label{rateproc3}
		\tau=\lambda_{n\to p+e^- +{\bar \nu}_e}^{-1}\simeq 887 \text{sec}\,.
	\end{equation}
	In (\ref{sumprocess}) one can safely neglect the contribution (\ref{rateproc3}) (during the BBN the neutron can be considered as a stable particle) \cite{bernstein}.
	Following \cite{bernstein} one can show that $\lambda_{e^+ +n\to p+{\bar \nu}_e}=\lambda_{n+\nu_e\to p+e^-}$.
	%
	%
	Inserting these results into  (\ref{sumprocess}) and (\ref{LambdaA}), one infers
	\begin{equation}\label{LambdafinA}
		\Lambda({T})\simeq 2\lambda_{np}=4\lambda_{e^+ +n\to p+{\bar \nu}_e}\,,
	\end{equation}
	which yields (using (\ref{ne-pnu-fin}))   \begin{equation}\label{LambdafinApp}
		\Lambda({\cal T}) =4 A\, {T}^3(4! { T}^2+2\times 3! {\cal Q}{T}+2!
		{\cal Q}^2)\,.
	\end{equation}
	
	\section{Linear stability analysis of ansatz (\ref{ansatz})}\label{C}
	Given that ansatz (\ref{ansatz}) plays a key role in the description of the BBN and the gravitational baryogenesis so it is essential to investigate whether it is an attractor solution or not. In the language of dynamical systems theory, attractor address situations where a collection of points in phase-space evolve within a given region, without leaving it. In other words, these points are stable in phase-space because them behave as sink or spiral sink. So, the advantage of an attractor solution is that it does not suffer from a fine-tuning of the initial conditions.
	
    To do so, putting ansatz (\ref{ansatz}) in the form $B=b a^{2\beta}$, together with introducing new variables $X_1=a$, and $X_2=\dot{a}$ in (\ref{19}), we reduce this second order dynamic equation to the following first order, consist of a autonomous system of differential equations 
	\begin{eqnarray}\label{Dy}
			\dot{X}_1&=&X_2=\mathcal{F}_1(X_1,X_2)~, \nonumber \\
			\dot{X}_2&=&\frac{2 X_2^2}{(\frac{\chi}{\xi} +1) X_1^{4 \beta +1}+2 X_1}-\frac{g_1 X_1^{4 \beta -2}}{(\frac{\chi}{\xi} +1) X_1^{4\beta +1}+2 X_1} \nonumber \\
			&&+\frac{g_2  X_1^{4 \beta }X_2^2}{(\frac{\chi}{\xi} +1) X_1^{4 \beta +1}+2 X_1}-\frac{4 }{3 (\frac{\chi}{\xi} +1)
				X_1^{4 \beta +3}+6 X_1^3}=\nonumber \\
			&&\mathcal{F}_2(X_1,X_2),
	\end{eqnarray}
where
\begin{subequations}
\begin{align}\label{g12}
g_1 =& \frac{\xi b^2 (\beta  (24 \frac{\chi}{\xi} +14)+7\frac{\chi}{\xi} +9)}{4 \beta }, \displaybreak[0]\\[1ex]
g_2 =&  \xi b^2 \left(48 \beta ^2 (2 \frac{\chi}{\xi} +1)-\beta  (36 \frac{\chi}{\xi} +14)-2(\frac{\chi}{\xi} +1)\right). 
\end{align}
\end{subequations}
Note that to derive of equations above, we have set $\omega=1/3$, and $\alpha=1/2$ together with $a_0=\tilde{M}=\kappa=1$. Now by serving the Jacobian matrix for the autonomous system (\ref{Dy})
\begin{eqnarray}\label{e3-6}
J\bigg(\mathcal{F}_1(X_1,X_2),\mathcal{F}_2(X_1,X_2)\bigg)=\left(
\begin{array}{cc}
\frac{\partial \mathcal{F}_1}{\partial X_1} & \frac{\partial \mathcal{F}_1}{\partial X_2}\\
\frac{\partial \mathcal{F}_2}{\partial X_1} & \frac{\partial \mathcal{F}_2}{\partial X_2}\\
\end{array}
\right)
\end{eqnarray}
we can say whether the solution (\ref{ansatz}) within phase-space $(X_1,X_2)$ can be an attractor or not. More precisely, the Jacobian matrix (\ref{e3-6}) is stable, indicating the solution (\ref{ansatz}) is an attractor provided that its trace and determinant i.e.,
\begin{align}\label{TD}
&tr=\frac{\partial \mathcal{F}_2}{\partial X_2},~~~\mbox{and}~~~ det=- \frac{\partial \mathcal{F}_2}{\partial X_1},~~~\mbox{where}~~~ \nonumber\\
&\frac{\partial \mathcal{F}_1}{\partial X_1}=0, ~~~~\frac{\partial \mathcal{F}_1}{\partial X_2}=1
		\end{align}
are negative and positive, respectively \cite{Coley}. By deriving $\frac{\partial \mathcal{F}_2}{\partial X_1}$, and $\frac{\partial \mathcal{F}_2}{\partial X_2}$, after some straightforward algebraic calculations, one can show that for $\frac{\chi}{\xi}<0$, and $\beta<0$, we have $tr<0$, and $det>0$, meaning that ansatz (\ref{ansatz}), enjoys stability and address an attractor solution.


\end{document}